# Echoes of the boom in hydraulic fracturing:

# What changes in Louisiana's energy complex reveal about global energy cycles

Alexander S. Kolker[1*], Robert Habans[2] Trey Cowan[3]

[1] Coastal Climates Institute, New Orleans, LA, alex@coastalclimates.org

[2] The Data Center, New Orleans, LA, robby@rhabans.com

[3] Institute for Energy Economics and Financial Analysis, Houston, TX. tcowan@ieefa.org

* Corresponding Author

alex@coastalclimates.org

## Abstract

The expansion of hydraulic fracturing (i.e. fracking) led to global shifts in energy markets that were felt acutely along the US Gulf Coast and particularly in the state of Louisiana. This article presents a data-focused examination of trends in the energy industry, and shows how they led to potentially counterintuitive, negative impacts in Louisiana's economy. Most notably, hydraulic fracturing led to a massive expansion in oil and gas supplies that contributed to a significant decline in oil prices in late 2014, contributing to significant job losses. Louisiana's economic downturns were tied to global shifts in energy markets that had major geostrategic implications. More broadly, results indicate that large and long period cycles in the supply, demand and price of oil can have widespread impacts on job creation and loss. In states like Louisiana with a large and diversified energy complex, the impacts of these trends can be highly uneven. One impact in the 21st Century appears to be highly concentrated areas of industrial development within a broader pattern of regional economic decline. Furthermore, trends identified in this study suggest the potential for the emergence of major cycles in the gas market. Whereas natural gas was previously mostly transported by pipelines and sold through long-term contracts, the emergence of liquified natural gas creates the potential for more volatile gas market dynamics. The implications of gas market cycles, while not yet known, could be of major economic significance.

## Introduction

The development of hydraulic fracturing in the United States released vast quantities of oil and gas, which led to profound shifts in global energy dynamics, geopolitics, economics and climate [1,2]. While the United States (US) had long been an energy importer, as a result of hydraulic fracturing in the 21st century, the US became one of the world's largest energy producers and exporters. The US Gulf Coast and, particularly, the state of Louisiana were at the center of this shift. Hydraulic fracturing, and the energy supplies it released, along with long-term declines in oil production shifted Louisiana's energy economy from one with strong links to production to one with strong links to manufacturing and export. Central to this shift was the development of large industrial projects, many valued in the billions of dollars.

While numerous studies have explored the economic impact of hydraulic fracturing, most regional studies focused on extraction zones. Few have addressed the impacts on exporting and processing zones and areas with complex links to multiple sectors of the fossil energy economy [3–5]. In this paper we ask, "how did Louisiana's economy change, in space and time, during the expansion in hydraulic fracturing?" This is not simply a question of how one process impacted one medium-sized state. Instead, it shines a light into the question of how capital, labor, and technology are distributed when there are major shifts in the production and distribution of some of modern society's most coveted natural resources.

Hydraulic fracturing produced large supplies of oil and natural gas, from fields in Louisiana and across the country, and the delivery of this material to Louisiana prompted a wave of investment concentrated in select corridors of the state. One might have expected this to lead to economic growth and employment gains. However, the increase in US oil supplies in the 2000s and 2010s was so abundant it contributed to global decline in oil prices in 2014 (Fig 1). This led to job losses that contributed to long-term patterns of economic decline in key regions of Louisiana.

**Fig 1. Price of oil and natural gas.** (A) Price of oil per barrel, normalized to 2024 US Dollars. (B) Price of natural gas in terms of dollars per BTU, including the price of LNG shipped to Japan (a common international metric for LNG), and the price of pipeline gas measured at the US Henry Hub (a standard metric for domestic US pricing). Data source: [6]

The 2014 oil price decline can be viewed in the context of multi-decadal cycles in the supply, demand, and price of energy that impact employment in the US Gulf Coast region and across the world. These are sometimes termed "super cycles." The post-World War II economic expansion, coupled with the oil price spikes of the 1970s contributed to employment gains in the Gulf Coast region, while the oil price declines of the mid-1980s led to significant economic contractions and job losses [7,8]. The rise of offshore drilling in the 2000s and early 2010s produced economic booms in south Louisiana. This paper will show that a new economic pattern developed in response to the 2014 decline in oil prices [9], where resource abundance contributes to widespread job losses and economic contraction contrast with intensive but concentrated capital investment in downstream energy uses.

Hydraulic fracturing emerged as a primary driver of 21st century global energy economics. While macroeconomic cycles, geopolitical conflicts, and acute environmental disruptions also shaped the energy landscape, the volume of hydrocarbons extracted by US shale production represented a structural supply shock large enough to realign international markets. For instance, the collapse in global oil prices following the supply surge directly prompted the 2016 formation of OPEC+, as traditional producing nations sought to coordinate output against rising American competition [10]. Within the shifting global stage, Louisiana and Texas occupied a pivotal position as primary nodes for extraction, transport, and downstream industrial processing.

This paper takes a time-series and geospatial approach, using datasets on energy production, employment, waterborne commerce, and economic growth. After review of hydraulic fracturing's background and Louisiana's distinctive energy complex, we examine major trends in oil and gas production and export in Louisiana and other significant regions in the United States and globally. We trace the current super cycle through trends in production, sales and state public revenues in Louisiana. The paper further describes impacts to employment, GDP, construction, waterborne commerce, and offshore activity. Finally, the discussion synthesizes these findings, highlighting how unprecedented industrial investment coexisted with lagging statewide growth and stark divergences across regional economies and industries.

# Background

## The hydraulic fracking boom

Hydraulic fracturing, colloquially known as fracking, is a technology used to extract hydrocarbons. It uses force to fracture sedimentary rocks and then injects water, sand (and sometimes chemicals) into them, allowing hydrocarbons to flow more readily [11]. The process is often used with other modern extraction technologies such as horizontal drilling. Shale rocks in particular have long been known to contain large quantities of hydrocarbons, but until the mid-2000s extraction was largely considered too difficult or too expensive [12]. Starting in the mid-to-late 2000s, these technological advances, coupled with favorable financial conditions and regulations, led to an expansion in hydraulic fracturing in the United States [12–14].

Hydraulic fracturing led to significant production increases in oil and gas in several parts of the United States (Figs 2-4), including the Haynesville Shale (northwest Louisiana and east Texas), the Permian Basin (Texas and New Mexico), the Eagle Ford Shale (Texas), the Bakken Formation (North Dakota, Montana, and parts of Canada) and the Marcellus Formation (Ohio, Pennsylvania, and West Virginia) [15]. Expanded oil and gas production in other parts of the world, including Canada, Australia, and Argentina is also linked hydraulic fracturing [16–19].

**Fig 2. Regional and national maps.** (A) Outline of the Gulf South region, noting the major states and cities. (B) Close-in view of south Louisiana, noting the location of major LNG facilities, and key parishes that are highlighted. (C) Shale plays (areas of extraction) in the lower 48 states. Source: (A) and (B) outlines are US Census TIGER files. (C) [15]

**Fig 3. Crude oil production in Louisiana and other key Gulf areas**. (A) Louisiana production 1981-2024. (B) Production in Louisiana, Texas, the Federal Offshore Gulf of America, and the USA as a whole. (C) Percentage of US production in Texas, Louisiana and the Federal Offshore Gulf of America. (D) US Crude Oil export 1981-2023, and those with oil exports from PADD 3 – the district that includes Louisiana and Texas. Source: [20,21]

**Fig 4. Natural gas production in the United States.** (A) Volume produced in Texas, Louisiana, and the Federal Offshore Gulf of America. Note that after 2001, natural gas from the Federal Offshore Gulf was removed from Louisiana and Texas and placed in its own separate category. (B) The percentage of these 3 states as a fraction of the USA's total production. (C) Production in several other states with large increases that are linked to hydraulic fracturing. *Source* [22].

US production increases shifted energy supplies and markets [6,12]. The United States became the world's largest producer and exporter of natural gas (Fig 5) and the world's largest producer of crude oil [6]. The US share of global oil production increased from about 8% in 2000 to nearly 16% in the 2020s [6]. For perspective, production growth in Texas, the state with the largest increase, was 4.4 million barrels per day (bpd) between 2009 and 2023, matching or exceeding the total production of many other globally significant producers such as Iran, Iraq, and the United Arab Emirates (Fig 5). In 2015 in response to increased domestic supplies, the US federal government lifted a 1975 ban on crude oil exports that originated during the energy crisis in the early 1970s [23]. Crude oil exports from the US increased rapidly after the export ban was lifted, reaching nearly 4 million bpd by 2023, making the US the 4th largest exporter of crude oil globally [6].

***Fig 5. Crude oil production in some of the world's largest producers****. (A) Volumes of crude produced in the world's top 8 producers. (B) Crude oil production as a percentage of global production for the world's top three producers. Source* [6]*.*

## Super-cycles in oil prices and global economic context

Changes in oil and gas supply due to hydraulic fracturing can be viewed in the context of long-term shifts in the supply, demand, and price of energy. Over the past 6 decades, large, long-term swings in the price of crude oil, often termed, "super-cycles," have greatly impacted the global economy and balance of power [7,24]. Table 1 describes key global and domestic events in super cycles, and Figure 6 shows the link between global oil price, number of rigs producing oil, and US employment in oil and gas extraction. The US Gulf Coast, as the largest energy complex in the world's largest economy, played a significant role in these cycles.

Table 1: Super cycles in oil price and key events along the Gulf Coast and across the globe.

| Years | Annual Oil Price ($2024/Barrel) | Key US Gulf Coast Events | Key Global Events |
|---|---|---|---|
| 1970 - 1985 | $14 - $140 | *Oil Boom<br>*Expansion of in-state drilling<br>*Job growth in Louisiana exceeds national average | *OPEC oil embargoes<br>*Iranian hostage crisis<br>*USSR invasion of Afghanistan<br>*End of Vietnam war<br>*Beginning of the petrodollar system |
| 1986 - 2000 | $24 - $80 | *Oil price crash<br>*Job losses in Louisiana and Texas<br>* Louisiana in state, land-based oil production begins to decline | *Cold war thaw/collapse of Soviet Union<br>*Corporate Reengineering.<br>*Era of globalization |
| 2001 - 2014 | $43 - $155 | *Expansion of offshore drilling<br>*Beginning of hydraulic fracturing<br>*Louisiana industrial construction boom<br>*Hurricanes Katrina and Rita | *September 11, 2001 attacks.<br>*China admitted to WTO, rapid expansion of Chinese industry<br>*Wars in Iraq and Afghanistan |
| 2015 - 2021 | $50 - $89 | *Oil price crash of 2014<br>*Reduction in offshore drilling<br>*Job losses in Louisiana<br>*Economic downturns in Louisiana<br>*Continued expansion of drilling in the Permian Basin<br>* Expansion of LNG exporting facilities | *COVID pandemic.<br>*Formation of OPEC+ in response to low oil prices<br>*Russia/Saudi Oil price war. |
| 2022 - 2025 | $81 - $109 | *Continuation of LNG exporting facilities and petrochemical buildout<br>*Emergence of data center as large energy users | *Russia Ukraine War<br>*Renewed Middle East conflicts |

The modern era of super cycles began in the early 1970s and coincided with the collapse of the Bretton Woods post-World War 2 financial system and OPEC-orchestrated oil embargoes [6]. Oil prices rapidly increased in the early 1970s, contributing to a decade of economic stagnation and decline in many industrialized countries[7]. In contrast, many oil-producing countries benefited from high oil prices. For example, real GDP in Saudi Arabia nearly tripled in the 1970s [25] . In the 1980s, the US Federal Reserve raised interest rates to control inflation, there were fewer tensions in the Middle East, and oil prices fell. Oil prices declined slowly in the early 1980s and more rapidly between 1984 and 1985 (Fig 1).

***Fig 6. The global link between the price of oil, the global rig count, and oil and gas jobs in the United States.*** *Note that the three fluctuate with approximately the same wave-like pattern, i.e. the super cycle. The global rig count is sourced from* [26] *the oil price data is sourced* [6]*, using their method for determining prices in 2024 US Dollars, and the number of workers is derived from the US Quarterly Census of Employment and Wages.*

In the US Gulf Coast region, the impacts of oil price swings on job creation were out of phase relative to the rest of the US. The 1970s were an era of job growth and economic expansion in Louisiana and Texas, while the 1980s were an era of economic contraction and job losses, particularly in drilling-related sectors [27]. This is the opposite of the rest of the United States, where the 1970s were dominated by slow or negative growth and high inflation, and the 1980s were dominated by relatively high job growth and low inflation. There were global impacts of the decline in oil prices in the 1980s. It contributed to the collapse of the Soviet Union and its client states, as the Soviet economy relied heavily on revenue from oil and gas exports [28,29].

The early 2000s marked a shift to the next super cycle phase as oil prices rose. In the broader Middle East, there were multiple conflicts and wars. The US Federal Reserve lowered interest rates, which often increases oil prices because low interest rates can reduce the US dollar's value, and oil is typically priced in US dollars. In the 2000s, rapid industrialization in China and other developing countries increased energy demand [6,12]. China's oil refining output more than tripled between 2000 and 2023, from about 4 million bpd in 2000 to over 15 million bpd, an increase that approximates the daily output of Saudi Arabia (11 million bpd) [6].

High energy prices and low interest rates in the early and mid-2000s created favorable economic conditions for investment in extraction via hydraulic fracturing, which set up the next super cycle phase in the late 2000s [12,13]. Oil and gas production in the US rapidly increased in 2008 (Fig 3). In Louisiana there was increased drilling on land in the Haynesville Shale, which utilized hydraulic fracturing, and in the offshore Gulf, which did not. Overall, US supply increases were so substantial that they contributed to the oil price crash of 2014 and 2015. In Louisiana, low oil prices led to job losses that are a focus of this paper [9]. But Louisiana was not alone; job losses occurred in other oil-producing states, including Ohio, Pennsylvania, Texas, and North Dakota [9]. Globally, the decline in oil prices was so significant that 10 countries joined together to form OPEC+, a companion to the OPEC cartel [10,30].

## Louisiana’s energy complex

Louisiana’s central role in global energy markets stems from its large and multifaceted "energy complex," a term used here to describe the broad swath of distinct but interconnected industries that extract, process, and transport oil and gas [31,32]. The roots of Louisiana's energy complex extend back to the early 20th century, when oil was discovered in 1901 in Jennings, Louisiana and nearby Beaumont, Texas. Louisiana's first large refinery was established in 1909 near Baton Rouge [33].

Today, Louisiana's energy complex includes operations that produce oil and gas from the state or nearby federal waters, and industrial facilities that refine oil and gas for fuels, and that use oil and gas as feedstock for chemicals, fertilizers, and plastics [32]. Energy generated with oil and gas is used to process metals and power manufacturing facilities. Liquified natural gas and crude oil are exported from ports in Louisiana and Texas to international markets [32]. Large data centers and spaceports are under construction that plan to use natural gas from regional supplies and pipelines [38]. For context on the scale of Louisiana's energy complex, the state hosts over 15% of the US' oil refining capacity, the second largest after Texas. Louisiana is the US's 3rd largest natural gas consumer. It is also the largest LNG exporting state in the US, which is the world's largest LNG exporting country [32].

Louisiana's energy complex is linked to its geographic position in the Gulf of America and at the mouth of the Mississippi River [32]. Louisiana's coast and adjacent offshore federal waters produce significant quantities of oil and gas. Since historically much of the US's oil supplies were sourced from international markets, Louisiana's deep ports along the Mississippi River, in Lake Charles, and offshore were a suitable receiving zone. Nearby oil refineries and petrochemical facilities could receive this oil, process it, and ship it across the US using the region's waterways [32]. Key waterways include the Mississippi River, which connects much of the central United States, and the Gulf Intracoastal Waterway (GIWW), which connects multiple Gulf states and the Mississippi River.

Additional factors also play an important role in Louisiana’s energy complex. Louisiana has six class 1 railroads (the highest capacity), which is one of the densest concentrations in the US [42]. The state offers an extensive pipeline network for hydrocarbons that includes Henry Hub – a major natural gas distribution center [43]. The country's Strategic Petroleum Reserve is located in salt domes along the coast of Louisiana and Texas; stored oil can be transported with ease to nearby refineries by pipeline [44]. The US Gulf Coast is also relatively protected from an energy security perspective; it would be difficult for a foreign military to attack because it is surrounded by US territory on multiple sides. Some authors also hold that Louisiana's regulatory framework is favorable to the energy complex, further encouraging industrial activities [45].

Hydraulic fracturing occupies a central role in multiple parts of Louisiana's modern energy complex. The Haynesville Shale in northwest Louisiana was one of the country's first large natural gas plays to utilize hydraulic fracturing [27,46]. Additionally, as hydraulic fracturing increased oil and gas supplies, supply was shipped to the Gulf Coast. Existing facilities, including the region's refineries and petrochemical plants, handled the extracted materials while other facilities were constructed or expanded to process the influx. New facilities include chemical manufacturing plants, iron smelters, and LNG exporting facilities. Industrial spending in Louisiana exceeded $100 billion between 2010 and 2023 [47], and numerous multibillion-dollar projects, including data centers and spaceports have been announced since the 2023 tally [48–50].

Although Louisiana's geography is particularly conducive to the energy complex, other locations have similar features. Texas has many similar characteristics as Louisiana, and the Gulf Intracoastal Waterway connected ports across the Gulf Coast to the Mississippi River. Globally, 75% to 80% of the world's crude oil trade is maritime [51], and processing it at the port of entry maximizes efficiency. Across the world, numerous oil refineries and industrial facilities are located near coasts and waterways. Large coastal energy complexes exist in Rotterdam in the Netherlands, China's Pearl River Delta, and the Persian Gulf [22,50].

From a regional economic perspective, Louisiana's natural resource and manufacturing industries represent classic "basic" or "tradable" sectors, as they produce goods and services for external markets. For example, Louisiana's oil refineries produce gasoline that is used nationwide. The economic footprint of the core industries that support the energy complex extends through deep supply chains and local consumption spillovers, with activities ranging from industrial maintenance to information technology and financial services to food services and retail [22,52]. In Louisiana, the industrial complex is tied to the state's maritime and water transportation infrastructure. Waterborne commerce and marine support activities are essential, and often labor intensive, extensions of the energy sector. Offshore drilling requires many people to build, service, and operate vessels, rigs, and heavy machinery [47–49]. Louisiana's waterways are among the most heavily used in the country, and a large fraction of the material carried on these waterways is energy related [50].

The employment footprint of the energy complex on job creation can be assessed by the "multiplier effect," which refers to the multiple of indirect jobs created by one direct job. Methodologically, the estimates stem from static input-output models that use data derived from the past to make implications about future job creation [51–53]. With a concentrated presence in Louisiana, high relative wages, and specialized regional supply networks, oil and gas industries tend to have large multipliers. The significant multiplier effect for oil and gas jobs produces appreciable spillover employment gains when the industries are growing and considerable job losses when industries are contracting. Industry fluctuations can have a widespread, but highly uneven impact on regional employment and GDP trends.

Here, we trace the uneven impacts of hydraulic fracturing and shifting global energy prices across specific industries and regional economies within the state. Hydraulic fracturing produced massive supplies of oil and gas that were transported to Louisiana, given its large energy complex. This led to expansive, though sometimes short-term construction jobs, but also contributed to long-term economic declines in drilling, and particularly offshore drilling. Overall, the pattern was broad-scale economic decline, despite pockets of growth.

## Methods

We examined trends in energy production, trade, and Louisiana's employment and gross domestic product (GDP) using publicly available data. Data are presented for multiple topics, including oil and gas production and export, jobs, gross domestic product, waterborne commerce, and tax revenue. While the timeframes of the available data vary to some extent between sources, most datasets start between the 1980s and early 2000s and extend through the early 2020s. A list of data sources and the topics they pertain to is presented in Table 2.

Table 2: Data Sources Used in This Manuscript

| Quantity | Resource/Agency | Website |
|---|---|---|
| Number of oil and gas rigs | Baker Hughes Rig Count | https://rigcount.bakerhughes.com/ |
| State-Level GDP trends | Bureau of Economic Analysis - US Department of Commerce | https://www.bea.gov/data/gdp/gdp-state |
| Tax abatement statistics for Louisiana | Good Jobs First - Subsidy Tracker | https://subsidytracker.goodjobsfirst.org/state/LA |
| Louisiana revenue from oil and gas royalties and severance taxes | Louisiana Department of Energy and Natural Resources (Now known as Louisiana Department of Conservation) | https://www.denr.louisiana.gov/page/revenue |
| Louisiana revenue from the Gulf of Mexico Energy Security Act | Office of Natural Resources Revenue, US Department of Interior | https://revenuedata.doi.gov/how-revenue-works/gomesa/ |
| Employment in Louisiana and Texas | US Bureau of Labor Statistics | https://data.bls.gov/PDQWeb/en<br>LA - Statewide, Private, Total<br>TX - Statewide, Private, Total<br>(settings for Query) |
| Employment Numbers for Louisiana and Texas - Chemical plants | US Bureau of Labor Statistics | https://data.bls.gov/PDQWeb/en<br>Similar but use NAICS 3251 |
| Total employment by state and metropolitan area | Bureau of Labor Statistics – Current Employment Statistics – State and Metro Area (BLS-SAE) | https://www.bls.gov/sae/ |
| Employment detailed by industry | Bureau of Labor Statistics – Quarterly Census of Employment and Wages (QCEW) | https://www.bls.gov/cew/ |
| Oil production in United States | US Energy Information Agency | https://www.eia.gov/dnav/pet/pet_crd_crpdn_adc_mbblpd_a.htm |
| Oil Exports from the United States | US Energy Information Agency | https://www.eia.gov/dnav/pet/pet_move_exp_dc_NUS-Z00_mbbl_m.htm |
| Dry Natural Gas Production in United States | US Energy Information Agency | https://www.eia.gov/dnav/ng/ng_prod_sum_a_EPG0_FPD_mmcf_m.htm |
| United States Natural Gas exports | US Energy Information Agency | https://www.eia.gov/dnav/ng/ng_move_state_a_EPG0_IM0_Mmcf_a.htm |
| Gas production globally by country | Statistical Review of World Energy | https://www.energyinst.org/statistical-review |
| Oil production globally by country | Statistical Review of World Energy | https://www.energyinst.org/statistical-review |
| Price of oil and natural gas | Statistical Review of World Energy | https://www.energyinst.org/statistical-review |
| Waterborne commerce trends | Waterborne Commerce Statistics Center | https://www.iwr.usace.army.mil/About/Technical-Centers/WCSC-Waterborne-Commerce-Statistics-Center/ |

While most data are presented directly as they appear in the original sources, in some instances we performed calculations or normalization the data. For example, the value of oil produced in the state was calculated by multiplying the amount extracted by the price, in 2024 US dollars -- as determined by the Statistical Review of World Energy (Fig 1).

When data on royalties and severance taxes are presented, we note most royalties and severance taxes are levied on oil and gas, but a small amount (always < 10% and often < 2%) are earnings from other minerals or plant liquids. Furthermore, the total value of mineral revenue in the state was calculated by summing royalties, severance taxes, and Gulf of Mexico Energy Security Act (GOMESA) funding. Detailed accounting of the other sources is available in the supplemental data section.

For economic trends, the predominant focus is statewide trends, though trends for specific regions within the state, as well as specific industry sectors and categories of trade goods are included to highlight patterns of divergence that coincide with the fracking super cycle. For employment, the focus is on upstream industries (oil and gas extraction, support activities for mining), downstream industries (petroleum and coal products manufacturing, chemical manufacturing), and various industries related to waterborne trade and construction. Table 3 presents a list of key jobs, along with their industry sector, and the data coding (NAICS) that is typically used to define these jobs.

**Table 3 –Oil and Gas Jobs within the Energy Complex**

| Category | Subcategory | NAICS | Industry |
|---|---|---|---|
| Direct oil and gas | Upstream | 211 | Oil and gas extraction |
| Direct oil and gas | Upstream | 213 | Support activities for mining |
| Direct oil and gas | Downstream | 324 | Petroleum and coal products manufacturing |
| Direct oil and gas | Midstream | 486 | Pipeline transportation |
| Chemical | Downstream | 325 | Chemical manufacturing |
| Chemical | Downstream | 326 | Plastics and rubber products manufacturing |
| Non-residential Construction | Sensitive industries | 2362, 237, 238 | Non-residential construction, Heavy and Civil Engineering Construction, Specialty Trade Contractors |
| Relevant water transportation | Sensitive industries | 483, 4883 | Water transportation, Support activities for water transportation |

To emphasize Louisiana's concentration of the jobs in the energy complex, we report the location quotient (LQ; Table 4), a common measure of an industry's concentration in a state or local economy [54]. The LQ is the ratio of that industry's share of a local area's (or state's) total employment to the same industry's share of total national employment. A high LQ (substantially greater than 1.0) suggests that the local industry produces goods and services for trade outside the local economy. High employment LQs identify the specialized industries that distinguish and often drive a state or local economy.

Table 4

Location quotient for selected industries, 2023

| NAICS | Industry | Louisiana Employment | LQ | Texas Employment | LQ |
|---|---|---|---|---|---|
| 211 | Oil and gas extraction | 5577 | 3.9 | 61480 | 5.95 |
| 213 | Support activities for mining | 22817 | 6.58 | 136288 | 5.43 |
| 2362 | Nonresidential building construction | 17171 | 1.59 | 113140 | 1.44 |
| 237 | Heavy and civil engineering construction | 42270 | 3.09 | 175267 | 1.77 |
| 238 | Specialty trade contractors | 65517 | 1.05 | 460705 | 1.02 |
| 324 | Petroleum and coal products manufacturing | 10718 | 8.06 | 21123 | 2.19 |
| 325 | Chemical manufacturing | 28222 | 2.55 | 86132 | 1.08 |
| 326 | Plastics and rubber products manufacturing | 4820 | 0.53 | 44424 | 0.68 |
| 483 | Water transportation | 9617 | 11.55 | 3907 | 0.65 |
| 486 | Pipeline transportation | 2491 | 3.79 | 20580 | 4.33 |
| 4883 | Support activities for water transportation | 13895 | 11.84 | 14338 | 1.69 |

Location quotients for selected industries, 2024

| | | Louisiana | | Texas | |
|---|---|---|---|---|---|
| NAICS | Industry | Employment | LQ | Employment | LQ |
| 211 | Oil and gas extraction | 5560 | 3.7 | 67466 | 6.18 |
| 213 | Support activities for mining | 21007 | 6.18 | 134825 | 5.47 |
| 2362 | Nonresidential building construction | 17976 | 1.6 | 120786 | 1.49 |
| 237 | Heavy and civil engineering construction | 47395 | 3.34 | 186492 | 1.81 |
| 238 | Specialty trade contractors | 68953 | 1.08 | 475968 | 1.03 |
| 324 | Petroleum and coal products manufacturing | 10801 | 7.91 | 21464 | 2.17 |
| 325 | Chemical manufacturing | 28719 | 2.59 | 89550 | 1.11 |
| 326 | Plastics and rubber products manufacturing | 4815 | 0.54 | 44598 | 0.69 |
| 483 | Water transportation | 9904 | 11.78 | 4304 | 0.71 |
| 486 | Pipeline transportation | 2534 | 3.73 | 22487 | 4.56 |
| 4883 | Support activities for water transportation | 14024 | 11.64 | 14729 | 1.68 |
| | Source: QCEW | | | | |

# Results

## Production and related activities in Louisiana

### Louisiana in the context of US oil production

US oil production has changed significantly over the past 40 years, and the rise of fracking has profoundly impacted the volume and location of US oil and gas production since the early 2000s (Fig 3). We highlight three key trends in domestic oil production. First, domestic crude oil production expanded in the late 2000s, reversing years of decline. US oil production was almost 9 million bpd in the early 1980s and dropped to between 5 and 6 million bpd by the mid-2000s. Starting in about 2008, US oil production began to increase - largely due to hydraulic fracturing [6,55]. By 2023, production approached 13 million bpd.

Second, while multiple states with sizable shale fields including New Mexico, Oklahoma, North Dakota, and Pennsylvania witnessed significant production increases, Texas drove the recent national increase (Fig 3). Production in Texas was close to 2.5 million bpd in the early 1980s and declined to just over 1 million bpd in the early 2000s. It increased starting around 2008, and by 2023 oil production in Texas averaged 5.5 million bpd- a roughly 5-fold increase over two decades. Most of Texas' production increases came from the Permian Basin and Eagle Ford Shale, which both hold large quantities of oil and gas that are geologically favorable to hydraulic fracturing [22,55]. Texas accounted for about 30% of domestic oil production in the early 1980s, 20% in the early 2000s, and over 40% in the 2020s.

Third, long-term production trends in Louisiana have been defined by a long-term decline within state boundaries, and a shift to production in the adjacent Federal Offshore waters (Fig 3). Crude oil production in Louisiana declined by about 80% between 1981 and 2023, from roughly 500,000 bpd to 100,000 bpd. The state's share of US oil production dropped from about 6% in the early 1980s, to less than 1% by the 2020s. In contrast, Federal Offshore production rose from 800,000 bpd in the early 1980s to 1.7 million bpd in the early 2020s. Federal offshore oil production was about 10% of US production in the early 1980s, over 29% in 2009, and between 14 and 16% in the 2020s. As a technical note, oil and gas produced in the Gulf of America is attributed to Gulf states differently. Louisiana's waters extend 3 miles offshore, while Texas' waters extend about 9 miles offshore, and Federal Offshore waters start at the states' respective border limits [56]. Given the amount of oil buried underneath the inner continental shelf, this demarcation places considerable amounts of oil and gas near Louisiana in the Federal Offshore category.

US natural gas production also increased substantially as a result of hydraulic fracturing, often on a similar timeframe as oil production (Fig 4). Overall, US natural gas production ranged between 16 and 20 trillion $ft^3$ from the 1980s to the early 2000s. Starting in approximately 2008, production increased, reaching 37.8 trillion $ft^3$ by 2023, a near doubling over 15 years [6,57].

As with oil production, multiple states with large shale formations experienced increased production due to hydraulic fracturing. Of these, Texas had the largest increase [57]. In Texas, natural gas production almost doubled between 2000 and 2023, from roughly 5 trillion $ft^3$ in 2002 to nearly 9.75 trillion $ft^3$ in 2023 (Fig 4). Production growth was notably strong between 2016 and 2023, with particularly significant growth in the Permian Basin [47]. Louisiana's gas production expansion was significant, a result of the Haynesville Shale located in the state's northwest corner. Statewide production increased from 1.2 trillion $ft^3$ in 2005 to 4.3 trillion $ft^3$ in 2023. On the other hand, natural gas production in Federal Offshore waters generally declined in the 21st century, dropping from about 5 trillion $ft^3$ in 2001 to less than 700 billion $ft^3$ in 2023. Prior to 2001, natural gas production from Federal Offshore waters was included in Louisiana's statistical total. That production was moved to a separate category in 2001, resulting in an artifact in Figure 4. Since 2018, Louisiana and Texas have accounted for about 10% - 12% and 25% of total US natural gas production, respectively [6,58].

In response to increased supply, US exports of natural gas rapidly increased. Prior to 2010, the US exported a small amount of LNG, mostly to Japan (Fig 7). By 2024, the United States was the world's largest LNG exporter, over 90% of which was exported from Louisiana and Texas. Louisiana and Texas' LNG exports increased from 0 to 2.7 trillion $ft^3$ and from 0 to 1.3 trillion $ft^3$ between 2010 and 2024, respectively [55]. Pipeline exports to Canada and Mexico also increased during this timeframe (Fig 7).

**Fig 7. US natural gas exports**. (A) Natural gas exports by type, LNG vs pipeline gas. (B) LNG exports by terminal. Source [6,59].

## Trends in oil and gas activities: sales and revenue

The value of oil production in Louisiana, which was determined by multiplying the amount extracted by the inflation-adjusted price, shows two major signals (Fig 8). The primary signal is a long-term downward trend that reflects declining oil production, and at times, lower prices (Figs 1,3). Inflation-adjusted oil sales declined from approximately $25.7 billion in 1981 to $2.4 billion in 2024. The secondary signal is an increase between the early 2000s and the early 2010s, and a sharp decline after 2014. The increase is attributable to rising oil prices, despite declining production, while the post-2014 decline is due to both declining prices and declining production.

***Fig 8. The relationship between oil production,*** *(A) the price of oil (in 2024 dollars) and the total value of production within the state of Louisiana. (B) The value of LNG exported from Louisiana, as determined by the volume of exports and the price of LNG exported. Source:* [6,32]

The finances of LNG export are quite different from the finances of oil drilling. Incremental LNG sales increased from $432 million in 2015 to over $21.7 billion in 2022 (Fig 8). LNG export values (incorporating the price paid at the point of export) increased from $870 million in 2016 to $21.3 billion in 2023. High natural gas prices in 2022 were influenced by Russia's invasion of Ukraine and global efforts to source from non-Russian gas suppliers [61]. Overall, total oil and gas sales in Louisiana ranged between $5.1 billion and $37.3 billion from 2011 to 2022.

When adjusted for inflation, the total revenue to Louisiana's government (including severance taxes, royalties, and GOMESA) has declined significantly over time (Fig 9). Inflation-adjusted revenue ranged between $4 billion and $5 billion in the early 1980s but was closer to $1 billion in 2024. Inflation-adjusted revenue exhibited similar price fluctuations as non-inflation-adjusted royalties and severance taxes, including a spike in the mid to late 2000s, and a decline after 2014. Severance taxes and royalties reached a peak of $1.96 billion in 2008, when oil prices were high, and experienced a notable decline after the 2014 oil price crash. Between 2018 and 2023, Louisiana received between $83 million and $156 million from the Gulf of Mexico Energy Security Act, which the state assigned to its Coastal Restoration Trust Fund (Fig 9).

***Fig 9. Sources of mineral revenue to the State of Louisiana****. (A) Total mineral revenue, nominal and inflation-adjusted, (B) Nominal mineral revenue by source. Source:* [58]

The long-run decline in production led mineral revenues (severance, royalties, bonuses, and rental payments) to fall in proportion to total state revenue [50]. Prior to the 1980s oil bust, mineral revenues ranged between roughly 30% and 40% of the state budget. After the 1980s, the percent of Louisiana's state budget revenue linked to oil and gas was less than 20%, and less than 10% between 2014 and 2020.

## Industry and employment trends

### Concentration of oil and gas industry sectors in Louisiana

The energy complex has a significant footprint in the Louisiana economy, as illustrated by the percentage of economic output (GDP) and employment in industries with direct ties to oil and gas (Fig 10). While these activities are concentrated within Louisiana, changes in the degree of those concentrations and the size of the industries reflect how the fracking transition has unfolded. Figure 10 presents the proportion of GDP attributed to various economic sectors, as a percentage of total state GDP for Louisiana, with Texas and the United States as a whole for comparison.

***Fig 10. The share of nominal GDP for selected industries in Louisiana, Texas, and the United States.*** *Source: US Bureau of Economic Analysis, GDP by industry in current dollars (SAGDP2).*

Energy sector trends shifted around 2008, corresponding with the onset of the boom in hydraulic fracturing. In Louisiana, the fraction of GDP in chemical manufacturing has trended upwards since 2008 and has consistently been greater than Texas and the rest of the country.

Meanwhile, Louisiana’s fraction of GDP in the upstream oil and gas sector has trended downward since 2008 but has been greater than the United States as a whole. Construction’s share of Louisiana GDP was similar to Texas and above the national GDP in the first part of the 21st century, though Louisiana's construction sector as a fraction of state GDP has expanded since 2010. While this may reflect intense construction at large petrochemical facilities, given the way that construction is classified in these data, it is not entirely clear this is the case. The share of GDP in the petroleum and coal product manufacturing sector has generally been flat in Louisiana, though there is a noticeable spike between 2003 and 2008. The fraction of GDP in business and professional services in Louisiana has not grown appreciably, suggesting that the state lags in diversification to high-paying services sectors.

While GDP measures the total output of core oil and gas industries, employment indicators also track the cyclical ebbs and flows of industry activity. Of course, long-term changes in the level and mix of employment also reflect transitions within the energy complex, including the reallocation of employment from upstream to downstream industries and labor-saving automation that tends to coincide with capital investment. Additionally, they reflect the changing extent to which the energy complex acts as a direct source of jobs for Louisiana workers. Note that Louisiana's oil and gas economy encompasses natural resources, manufacturing, transportation, and other traded industries that tend to be highly specialized and concentrated in the state, as indicated by high employment LQs (Fig 12, Table 4).

In terms of direct oil and gas employment, upstream and downstream industries have followed different trajectories over the last decade. Figure 11 shows employment trends upstream and downstream oil and gas, several of which are identified in Table 3. In Louisiana, employment in upstream oil and gas extraction fell from 49,885 jobs in 2014 to 31,529 jobs in 2017. Employment stood at 26,567 jobs in 2024 – a decline of 47% since 2014 – and has remained under 30,000 since 2020. Between 2014 and 2024, employment also fell in Texas from 294,919 jobs to 202,291 jobs (31%). Upstream employment declined at the national level from 636,083 jobs to 395,610 jobs (38%) during the same period. Louisiana's upstream oil and gas extraction (LQ = 3.9) and support activities for mining industries (LQ = 6.6) are highly concentrated in Louisiana, along with other states with significant mining and oil and gas production (Fig 12, Table 4).

**Fig 11. Private employment in Louisiana in upstream (production) and downstream (refining, chemical manufacturing etc.) sectors of the oil and gas industry:** (A) Total employment. (B) Employment as a percentage of state total employment. Source: US Bureau of Labor Statistics, Quarterly Census of Employment and Wages (QCEW).

**Fig 12. Concentration of key parts of the oil and gas industry in Louisiana as compared to other states using the location quotient (LQ).** Source: US Bureau of Labor Statistics, Quarterly Census of Employment and Wages (QCEW).

Meanwhile, downstream industries have added jobs, particularly in chemical manufacturing, where Louisiana has outpaced national growth. Between 2010 and 2024, chemical manufacturing employment grew by 28% in Louisiana, from 22,498 jobs to 28,719 jobs, compared with an increase of 27% in Texas and 14% nationally. Chemical manufacturing's concentration in Louisiana (Fig 12, Table 4), measured as the location quotient (LQ), is the highest among states (LQ = 2.59), and it increased during the 2010s. Conversely, the other major downstream manufacturing industry, petroleum and coal products manufacturing (refining), has seen employment in Louisiana waver between 10,000 and 12,000 jobs since the 1990s. National employment in this sector decreased steadily over the same period, from 154,000 jobs in 1990 to 110,000 jobs in 2024. The industry's employment concentration in Louisiana is also the highest among states (LQ = 7.91).

The shifts in upstream and downstream industries in Louisiana are also reflected in the changing concentration of employment in the state relative to the nation. Figure 13 shows changes in LQ over time. Since roughly 2000, the concentration of industry employment in Louisiana relative to the national economy has fallen for oil and gas extraction and support activities (upstream), but it has increased for petroleum and coal products manufacturing and chemical manufacturing (downstream).

**Fig 13. The changing nature of the location quotient of jobs in Louisiana**. Note the downward trend in support activities for mining and the upward trend in petroleum and coal products manufacturing.

## Water transportation and port employment

Louisiana's water-based economy has multiple ties to the energy complex. It has the highest private employment concentration of any state in water transportation (Fig 14; NAICS 483, LQ = 11.8, Table 4) and support activities for water transportation (port and cargo activities and navigational services, NAICS 4883, LQ = 11.6). These jobs include workers who pilot and crew vessels that transport workers to offshore oil rigs, and ocean-going ships that import crude oil, and barges that distribute refined products across the country (Table 3). Together, water employment and water support activities make up about 1.5% of total Louisiana employment, which is on par with the amount of downstream oil and gas jobs in private employment in 2024 (Figs 11,14). Here we highlight two major examples of water-based energy jobs: offshore oil and gas production and the transport of energy and petrochemical products on Louisiana's waterways (Fig 15, 16).

**Fig 14. Employment in Louisiana in water transportation and support activities as a proportion of the total number of jobs in the state.** Source: US Bureau of Labor Statistics, Quarterly Census of Employment and Wages (QCEW)

**Fig 15. Employment in the Houma Thibodaux region in south-central Louisiana.** The region has long been considered a hub for jobs in the offshore oil and gas industry. Source: US Bureau of Labor Statistics, Quarterly Census of Employment and Wages (QCEW).

**Fig 16. Trends in waterborne commerce in selected along key waterways in Louisiana.** Trends are presented for the lower Mississippi River, the Gulf Intracoastal Waterway, and the Lake Charles Deep Draft. Source: [60]

Increased oil production in Federal Offshore waters, which started in the 1990s, is strongly associated with Louisiana. One particularly integral asset is Port Fourchon in Lafourche Parish, which is the primary offshore services port in the Gulf of America (Fig 15). Numerous jobs that support offshore activities are located in neighboring Terrebonne Parish. The number of offshore rigs in Louisiana peaked in the early 2010s; however, active rig count fell precipitously in the middle of the decade, concurrent with the 2014 oil price decline (Fig 17). The number of Louisiana offshore rigs rose between 2010 and 2014, which suggests regulations enacted after the BP/Deepwater Horizon oil spill did not significantly reduce offshore oil drilling [59]. Louisiana's land rig count also declined in the 2010s, though with a slight variation in timing from offshore rigs. In Texas, where most extraction is on land, large swings in the active rig count are consistent with the technology. Hydraulically fractured wells produce large volumes of oil, but only for a short period of time. Globally, the number of oil rigs follows wave-like alignment with the changes in oil price. (Fig 6).

**Fig 17. Number of drilling rigs in Louisiana, Texas, and the offshore regions of both states.** (A) Total number in all locations. (B) Total number in Louisiana. (C) Total number in Texas. (D) Number of land-based and water-based rigs in Louisiana. Source: [61,62]

Louisiana's ports and waterways also show a distinctive imprint of the state's energy economy. Table 5 presents a summary of shipping statistics for the Mississippi River, the Gulf Intracoastal Waterway, the Lake Charles Ship Channel, and the Calcasieu River and Passes. We categorize commodities into 5 major groups: Direct Energy, Chemicals, Metals, Wood and Agriculture, and Other. On the Mississippi River – which is the largest pathway for waterborne commerce in the United States – Direct Energy, Chemicals, and Metals comprised 56.8% of all commodities. Along the Gulf Intracoastal Waterway, the Lake Charles Ship Channel, and the Calcasieu River and Passes the fraction that is directly linked to energy is even higher.

**Table 5 - Waterborne Commerce in Key Louisiana Waterways**

Millions of Short Tons, Average 2019-2023

| | Mississippi River, Baton Rouge to Head of Passes | Gulf Intracoastal Waterway, Louisiana Portion | Lake Charles Deep Draft Channel | Calcasieu River and Passes |
|---|---|---|---|---|
| Total | 435.0 | 70.4 | 51.0 | 55.5 |
| Direct Energy | 153.4 | 40.8 | 34.0 | 47.9 |
| Chemicals | 61.2 | 13.1 | 9.8 | 4.5 |
| Metals | 32.2 | 4.1 | 2.7 | 0.3 |
| Wood and Agriculture | 162.7 | 1.3 | 0.4 | 0.3 |
| Other | 27.0 | 11.2 | 4.0 | 2.5 |

Percentage of Total Transport 2019-2023

| | Mississippi River, Baton Rouge to Head of Passes | Gulf Intracoastal Waterway, Louisiana Portion | Lake Charles Deep Draft Channel | Calcasieu River and Passes |
|---|---|---|---|---|
| Direct Energy | 35.3 | 57.9 | 66.7 | 86.3 |
| Chemicals | 14.1 | 18.6 | 19.3 | 8.1 |
| Metals | 7.4 | 5.8 | 5.4 | 0.5 |
| Wood and Agriculture | 37.4 | 1.8 | 0.8 | 0.6 |
| Other | 6.2 | 15.8 | 7.9 | 4.5 |

**Table 5 Summary statistics on waterborne commerce in key waterways in south Louisiana.** Data were grouped into five major categories, and the total volume in each category (measured in short tons) was summed. Direct energy includes items directly sourced from fossil energy (e.g., crude oil, coal, gasoline). Chemicals include products that are sourced from oil or gas (e.g., benzene and toluene), or are energy intensive to create (e.g., nitrogenous fertilizer). It also includes products that can come from oil and gas or other sources (e.g., sulfur). Metals include refined and raw metals (e.g., iron and steel, and aluminum ore). Wood and agriculture products include forestry products, (e.g., wood chips), and major crops (e.g., corn, wheat, rice), as well as smaller agricultural items. The other category includes an array of products not easily classified into the above categories. Data source: [48,63].

Water transportation employment more than doubled between the late 1990s and early 2010s (Fig 14). This was most likely tied to production in Federal Offshore waters, as most of the job growth was in Lafourche Parish, home to Port Fourchon. At the same time, the port-focused support activities industry was flat or declining in most parishes with significant port employment outside of Terrebonne and Lafourche. Employment in both industries fell with oil prices in 2014. Thus, the post 2014 decline of traffic is also reflected in the combined trends in water employment and its related support activities (Fig 16).

As shown in Figure 16, the total volume of material on the Mississippi River increased from 2002 to 2017, and then declined after 2018, with crude oil and gasoline leading the decline. This trend indicates that the United States imported less crude oil and refined less gasoline in Louisiana, as production increased in Texas due to hydraulic fracturing. Indeed, United States' energy imports have been declining since about 2008, and by 2019, the United States was a net energy exporter [3,64]. In contrast, along the Calcasieu River and Passes, the category "hydrocarbon & petroleum gases liquefied and gaseous" increased substantially after 2016, reflecting LNG exports.

## Construction boom

The energy complex has a distinct footprint in Louisiana’s construction sector. For example, heavy engineering and civil construction and nonresidential building construction are both concentrated in the state (NAICS 237, LQ = 3.09, and NAICS 2362, LQ = 1.44, respectively, Table 4, Fig 18). When looking at detailed construction industries, the largest concentrations are in categories with strong ties to large energy-related projects: oil and gas pipeline and related structures construction (NAICS 23712, LQ = 8.7), industrial building construction (NAICS 23621, LQ = 3.75), and other heavy and civil engineering construction (NAICS 2379, LQ = 8.63). While the category "Other heavy and civil engineering construction" includes construction related to coastal, water, and flood management – reflecting the state's efforts in coastal restoration and protection – it can also include energy-linked construction projects. This category is uniquely concentrated in Louisiana: no other state has an LQ exceeding 2.0 (Table 4).

**Fig 18. Employment in construction and employment in selected south Louisiana parishes**: (A) Urban areas. (B) Parishes in southwest Louisiana and the Mississippi River industrial corridor. Source: US Bureau of Labor Statistics, Quarterly Census of Employment and Wages (QCEW).

Construction has become increasingly central to the function of Louisiana’s energy complex as a job creator. Between 2008 and the present day, multiple large, capital-intensive industrial projects that use natural gas and oil – for energy, and as a chemical feedstock – were constructed in Louisiana. One report from Louisiana's Department of Economic Development cited $100 billion of new capital expenditure between 2016 and 2023, much of which was linked to natural gas and petrochemicals [61]. More recently, several large data centers in north

Louisiana that use natural gas-generated electricity are planned or under construction [62]. A large spaceport was recently announced that will use chilled methane – the major component of natural gas – as a rocket fuel [63].

Construction employment spiked during the 2010s, driven in part by oil and gas-related industrial facilities. Between 2007 (the peak just before the Great Recession) and 2018, employment in nonresidential construction grew by 17%, with the majority occurring between 2014 and 2018 (Fig 18). The LQs of "oil and gas pipeline and related structures construction" and "industrial building construction" increased temporarily in the 2010s. This suggests that construction in Louisiana played an important role in national construction trends that were also linked to hydraulic fracturing. In contrast, employment in all other industries grew by only 5 percent between 2007 and 2014, with slight declines noted between 2014 and 2018.

The construction boom was highly localized. Many jobs were in South Louisiana's two main industrial corridors: in southwest Louisiana near the city of Lake Charles (Calcasieu and Cameron parishes) and along the Mississippi River south of Baton Rouge (Ascension, Iberville, Plaquemines, St James, St. John the Baptist, and St. Charles parishes; Figure 18). Cameron Parish stands out as a particularly interesting case study. During a period when multiple large LNG exporting facilities were constructed in and near the parish (~2014 to 2020), employment in the parish underwent a steep peak and decline (Fig 18). At the 2018 peak, employment reached 12,500 jobs, more than double the parish's full-time population of 5,600 people reported in the 2020 census [64]. However, the post-peak low was slightly higher than the pre-peak low, likely reflecting the presence of more permanent manufacturing jobs.

## Slow, geographically uneven growth within Louisiana

### Parish employment and GDP

Moving shifts across industries to uneven trends at the local level, Figure 19 presents employment and GDP trends across all industries at the parish level. The map illustrates local variation in trends over two periods 2001-2023 and 2014-2023, with the latter capturing trends since the late 2014 oil price decline. Note that a mix of trends are reflected in these patterns, only some of which are linked to the energy complex. Since 2001, trends in the southeast and southwest corners of the state were directly affected by Hurricanes Katrina and Rita (2005, severely affecting Orleans and St. Bernard Parishes, among others), among the most destructive storms in United States history. Additionally, long-term employment growth has been modest in much of the state outside of suburban parishes, especially in southeast Louisiana near New Orleans and Baton Rouge (especially in Livingston, Ascension, St. Tammany, and Tangipahoa). The sharpest declines have been in rural parishes, especially in the northeast corner of the state.

**Fig 19. Maps of employment and GDP change in Louisiana's 64 parishes**. Shown here are data for two time periods, 2001-2014 (before the oil price decline) and 2014 to 2023 (after the oil price decline). (A) GDP growth. (B) Employment growth. Sources: (A) US Bureau of Economic Analysis, Real Gross Domestic Product:], Retrieved from FRED, Federal Reserve Bank of St. Louis; (B) Source: US Bureau of Labor Statistics, Quarterly Census of Employment and Wages (QCEW).

Since 2014, decline has been more widespread. In addition to the perennially struggling rural north of the state, job losses have been especially severe in coastal parishes, where most have lost jobs. These include Lafourche and Terrebonne Parishes, where GDP declined after 2014, leading to a double-digit, decade-long economic downturn in these two parishes. These parishes are notable because they served as a major focal port for jobs associated with offshore drilling in the Gulf of America. While the map shows employment and GDP decline in Cameron Parish, this masks the reality of volatility. Cameron Parish experienced a long-term GDP decline from 2001 to 2013, followed by an increase in GDP and jobs between 2013 and 2018, roughly aligned with the construction of several large LNG exporting terminals. The overall parish-level pattern is marked by areas of strong employment and GDP growth juxtaposed with widespread decline, punctuated by pockets of extreme volatility. The pattern is prevalent in coastal parishes, where much of the energy complex is located.

While long-term growth has been modest in much of the state, suburban parishes in southeast Louisiana near New Orleans and Baton Rouge (especially in Livingston, Ascension, St. Tammany, and Tangipahoa) sustained GDP growth. GDP declines were also significant in northeast Louisiana, which is part of the "Mississippi Delta," an agricultural area of former plantations and long been one of the poorest parts of the United States [65,66].

## Regional employment trends

Shifts in oil and gas production had uneven effects on regional economies across Louisiana. Figure 20 shows employment in Louisiana's largest metropolitan areas between 1990 and 2023. Regional employment trends in Louisiana tend to be somewhat volatile. As in other states, Louisiana’s regional economies reflect the national business cycle (recessions and recoveries). Conditions related to the energy complex also shape regional employment trends, as do localized impacts of disasters (*e.g.*, hurricanes, COVID-19).

**Fig 20. Non-farm employment for selected metropolitan statistical areas (MSAs) in Louisiana.** Source: US Bureau of Labor Statistics' State and Area Employment.

Amid a period of consistent national job growth from during the 2010s (the interval between the Great Recession and the pandemic), employment trends in most of Louisiana's regional economies have been marked by slow growth and, in some cases, volatility, especially since 2014 (Fig 20). As of 2023, Baton Rouge was the state's only metropolitan area to post substantial job gains after 2014 (7.5%), about half of which has occurred during a strong pandemic recovery, relative to other metro areas in the state. The Lafayette metro, where a significant portion of upstream oil and gas employment is concentrated, lost 7.4% of employment between 2014 and 2016. Lafayette's COVID recovery appears strong and employment in 2023 was only 2.8% below 2014 levels. Aside from the Great Recession (2008), the New Orleans metro saw several years of consistent growth as the city recovered from Hurricane Katrina in 2005. By the late 2010s, growth had leveled off, and employment was slightly below pre-pandemic levels as of 2025.

The most significant swings occurred in the Lake Charles and Houma-Thibodaux metro areas (Figs 15,20). Total employment in the Houma-Thibodaux region fell by 15.6% between 2014 and 2017, driven by declines in the manufacturing, oil and gas and pipeline transportation, and natural resources and mining sectors. All of the aforementioned sectors are intertwined with the energy complex, particularly the offshore oil and gas industry, which have had major hubs in the region. Lake Charles experienced a 29% increase in employment between 2011 and 2018, driven by oil and gas-related manufacturing and LNG construction. The spike tapered off before the pandemic, and employment in 2025 was roughly 8% higher than early 2010s levels.

The Lafayette metro replaced the Shreveport metro as the state's third-largest regional labor market in the late 2000s, and Shreveport has since been marked by a long-run, relatively consistent decline. Despite production from the Haynesville Shale, Shreveport employment did not show the same volatile swings after 2014 as in the state's other mid-size and smaller metropolitan regions: sharp drops in Lafayette and Houma-Thibodaux and a rapid increase in Lake Charles.

# Discussion

## Overview

To summarize, the research herein points to several key findings. First, hydraulic fracturing led to global changes in energy markets, which transformed the United States from the world's largest energy importer to one of the world's largest energy exporters. Though hydraulic fracturing only modestly increased gas production in Louisiana, nationally it generated vast supplies of hydrocarbons that were routed to the US Gulf Coast, realigning Louisiana's energy complex. The impact on state revenue, employment, economic stability, infrastructure investment, and trade flows was broad and multilayered. Second, economic gains and losses associated with hydraulic fracturing were not evenly distributed by industry or by geography. As Louisiana's position relative to the national and global energy economy evolved, energy-related investment concentrated in some regions while other areas stagnated. The shift from the upstream (*i.e.*, production/ extraction) to downstream and power sector (*i.e.*, chemical manufacturing and LNG export) restructured the energy complex. As investment concentrated in select regions and declined in others, spillover jobs shifted as well – leaving behind areas of stagnant or declining employment. Finally, Louisiana's energy complex is linked to global and long-term cycles in the production, distribution, and consumption of fossil energy. While the United States predominantly influenced the demand-side of energy economics in the past, the availability of new supplies in and around the Gulf Coast profoundly changed the dynamic. Overall, hydraulic fracturing produced a new distribution of winners and losers, in both time and space. Understanding how positions change should inform local and global conversations about the energy economy.

## Super Cycles and the Energy Transition

The shifts we document in Louisiana can be viewed as a crucible of the global "energy transition." The term "energy transition" is used broadly and can have different meanings to different authors [51,67]. For some, it implies a shift away from fossil energy toward renewable sources; for others, it refers to transition away from oil and toward natural gas. Trends in Louisiana are broadly consistent with the latter definition. The transition is not a temporary market fluctuation but rather a structural change driven by the long cycles of global investment in energy infrastructure and industrial capacity, with profound local consequences for the state of Louisiana, its workers, and its regional economies.

Since the 1970s, there have been major cycles in the price of oil, with each cycle lasting about 20 to 30 years (Fig 6, Table 1). The price cycles were linked to major geopolitical and economic events, including the oil embargoes in the 1970s, other production decisions by OPEC, the formation of OPEC+ in 2016, conflict in the Middle East, the collapse of the Soviet Union, the industrialization of China and other developing countries, environmental regulations, and interest rates. The US Gulf Coast, as the largest energy complex in the world's largest economy, has had a front-row seat to key elements of these cycles.

The current super-cycle phase is framed by the influence of hydraulic fracturing that generated increased oil and gas supplies starting in 2008, which led to an oil price decline in 2014. The United States became the world's largest oil producer, exceeding Saudi Arabia and Russia, which had previously dominated global production. The United States also became the world's largest LNG exporter, surpassing countries like Australia and Qatar. National production growth was closely linked to hydraulic fracturing in Texas' large land-based oil fields, notably the Permian Basin and the Eagle Ford Shale.

The period between 2014 and 2016 marked a seminal transition for Louisiana's energy economy and redefined its relationship with broader United States and global economic trends (Fig 6). In 2015, job growth in Louisiana began to decouple from national trends in job growth (Fig 21). For much of the post-World War II era, Louisiana's job trends generally mirrored national trends. However, job growth in Louisiana flattened during 2015, while the US job growth increased. As of December 2025, national employment had grown 11.5% since August 2015. Over the same period, Louisiana employment had only grown 0.6%. Louisiana had essentially no job growth between 2015 and the 2020 pandemic, while the United States grew 7% over the same period. Regionally, many Louisiana metropolitan areas experienced either slow or negative job growth between about 2008 and 2023 (Figs 19, 20). The early 1980s oil bust was the only other period since the postwar era when the state lagged behind the nation for a prolonged period (Fig 21).

**Fig 21. Employment in Louisiana, Texas and the United States, referenced to employment in 1950.** Source: US Bureau of Labor Statistics: State and Area Employment

The oil price crash contributed to large job losses and GDP declines in parts of Louisiana (Figs 19,21). Consequences were felt acutely in south Louisiana communities where employment was predominantly dependent on offshore oil and gas production, and rippled through the energy-adjacent, indirect workforce that relied on robust offshore oil and gas activity. The Houma-Thibodaux metropolitan region (Terrebonne and Lafourche parishes), anchored heavily to labor-intensive offshore drilling and engineering services, absorbed steep declines after the 2014 price collapse (Fig 15). By 2023, its GDP and employment remained 20% and 16% below the 2014 peak, respectively. Conversely, the Lake Charles region (Cameron and Calcasieu parishes) leveraged its deep-water infrastructure to attract immense downstream LNG export and industrial processing facility construction. Fueled by construction, employment in the Lake Charles region spiked by 30% between 2010 and 2018, only to experience a contraction thereafter, leaving net growth since 2010 at 8% as of 2024 (Figs 18,22). While 2014 also brought slowed growth to the larger regions of New Orleans and Baton Rouge, similar contrasts mark the Mississippi River industrial corridor between the two regions. In addition, as in-state production declined, and prices generally trended downward (Fig 9), the value of product produced in-state declined, as did revenue to state coffers, both of which likely took an additional economic tool on the state.

**Fig 22 Employment in the Southwest corner of Louisiana concentration in Cameron and Calcasieu parishes.** Source: US Bureau of Labor Statistics, Quarterly Census of Employment and Wages (QCEW).

National shifts in energy production alter the flow of energy in the Gulf, which also led to job losses in Louisiana. For example, crude petroleum and gasoline shipments along the Mississippi River began to decline in 2017 as internationally sourced crude was replaced with domestic crude – of which Texas was the largest source (Figs 3, 16). Crude oil imports declined in PADD 3, the designated energy region that encompasses the US Gulf Coast, from about 5.8 million bpd in 2004, to about 1.3 million bpd in 2024 [21]. In the early 2020s, several oil refineries along the Mississippi River in south Louisiana closed [69].

## Chains of decline, islands of growth

Research has documented the positive spillover benefits of hydraulic fracturing on regions new production developed [70,71]  In this paper, we explore the flipside of these consequences for Louisiana. New drilling shifted away from Louisiana and its offshore energy complex, and capital flowed to downstream processing and export infrastructure. While positive job spillover effects (*i.e*. increased indirect jobs) are linked to boom regions, Louisiana's bust has elicited corresponding negative effects through the decline in the upstream production supply chain. Industry-level trends and the state's lagging total employment growth suggest consequences of the negative spillovers due to lost production in Louisiana have outweighed the positive spillovers expected of industrial and infrastructure investment.

The decline in offshore drilling jobs that occurred after the 2014 oil price decline had wide-ranging impacts on the economy across south Louisiana. The changing mix of jobs tied to Louisiana's energy complex, as well as the gross number of jobs, is central to these differential impacts. Offshore operations often involve complex equipment, including vessels and aircraft that require deep, labor-intensive supply chains across multiple industries to operate, maintain, and service [71]. In contrast, investments in downstream industry create short-term, project-based construction jobs and are otherwise capital intensive. Additional factors very likely played a role, including increased efficiencies in drilling and information technology that contribute to increased productivity and a smaller workforce [72].

Post-2014 Louisiana reveals a distinct phase of the "resource curse"—one that moves beyond simple extraction dynamics [73]. Classically, a resource curse unfolds as a natural resources boom drives up costs for manufacturing, entrepreneurship, professional services, and other tradable sectors, leading to the paradoxical outcome in which a region blessed with resources experiences slower long-run growth and less diversification. Arguably, Louisiana’s oil and gas booms and busts through the 20th century resembled this archetype, as the state enjoyed periods of above-average growth but failed to sufficiently diversify its economy [74]. The oil price collapse of the mid-1980s resulted in job losses in the drilling sector, professional services, and other sectors that were indirectly sustained by Louisiana's upstream production [75].

In the 2010s, production shifted away from offshore drilling and its high productivity spillovers, though significant capital flowed into downstream processing and export infrastructure [76]. As natural gas became the major driver of capital investment, job creation, and regional agglomerative spillovers, Louisiana's energy complex evolved. Since 2014, the resource curse entered a new phase, one linked to supply abundance, in which an ongoing construction boom and modest job gains in downstream industries are paired with historically depressed conditions in the upstream industries. The resulting pattern of these economic transitions appears to be an economic islanding effect, and the creation of small regions of economic growth within larger regions of economic stagnation or decline.

While the 1980s oil price crash also significantly affected Texas, the sheer scale of the Texas economy and the presence of larger metropolitan areas (Houston, Dallas-Fort Worth, Austin) that host diverse sectors – finance, health care, and technology – appears to have mitigated the impact. The state’s economy was able to pivot into non-energy tradable sectors [77–79]. We suspect that Texas has more successfully leveraged energy wealth to fund broad-based human capital development, post-secondary education, and incubate other diversified clusters (*e.g.,* health care, technology). One notable example is the Permanent University Fund, which dedicates oil revenue from fields in West Texas to several of the state's large public universities [2,44,78]. Consequently, while Texas remains the nation's largest oil producer, the energy sector has shrunk as a percentage of its total GDP (Fig 10). When the 2014 shock hit, Texas's diversified economy absorbed the blow with fewer statewide impacts [76].

In Louisiana, economic gains and losses do not occur in isolation but interact with acute environmental and demographic vulnerabilities. In recent decades, the state's lagging economy has been compounded by disasters fueled by climate change, coastal land loss, and infrastructure deficiencies. Catastrophic storms – including Hurricanes Katrina and Rita (2005), Laura and Delta (2020), and Ida (2021) – inflicted widespread physical destruction across the state's primary industrial and maritime corridors and disrupted local property insurance markets [79]. These intersecting vulnerabilities have also contributed to slow or negative population growth across multiple south Louisiana parishes in a manner that is complex to disentangle from underlying economic drivers of migration and demographic change.

## Learning from the Fracking Boom: From Domestic to Global Markets

The structure of global energy markets is changing, and the US Gulf Coast is at the center of these changes. Although for most of the post-WWII era the United States was the largest source of global demand and the US Gulf Coast was a major importing center, since 2008 the United States has become a large driver of supply, and since 2016 the US Gulf Coast has become a major exporting center (Figs 3-5,7). Furthermore, energy super cycles were previously mostly confined to the oil market [80], but the current super cycle is linked to both oil and gas. Both of these changes are driven by the large quantities of US-sourced oil and gas that were produced by hydraulic fracturing.

The economic downturns associated with the 2014 oil price decline were hardly limited to Louisiana; instead, they were global and had major geopolitical ramifications. In response to chronically low oil prices and their negative impact on drilling-based economies, in 2016 ten oil-producing countries joined OPEC+, an OPEC-aligned cartel that coordinates with OPEC to control production, and thus prices. This increased cartel influence over oil markets from 38% to 59%, based on 2022 production statistics [80]. The volume of US-produced LNG was so significant that it replaced Russian gas in several key international markets and caused Russia's state-owned gas company Gazprom to delay or scale back several projects [80]. In inflation-adjusted 2025 US dollars, between 2014 and 2016, Russia's GDP declined $2.06 trillion to $1.28 trillion – the country's largest GDP decline since the collapse of the Soviet Union [81]. Saudi Arabia's GDP declined from $787 billion to $689 billion [2,44,82]. Viewed from this angle, the economic declines – and export booms – in south Louisiana appear part of a major geopolitical restructuring.

Looking to the future, it is critical to understand the influence of large, long-period cycles in the natural gas market could impact economies and communities [81]. Natural gas was historically confined to national and regional markets, which were less cyclical and less global, because it was typically sold through long-term contracts and moved about by pipelines [2,44,82]. The emergence of largescale LNG trade has allowed natural gas to be sold more widely across the world and has allowed the price to fluctuate on more of a spot market. New trading dynamics, combined with ample new supplies made available by hydraulic fracturing, create conditions where super-cycles in natural gas can emerge.

As natural gas, and particularly liquefied natural gas, becomes a globally traded commodity, market dynamics become significant to Louisiana. There are large interests that benefit from high gas prices (*e.g.*, producers in the Haynesville shale and potentially LNG exporters), as well as large interests that benefit from low gas prices (chemical manufacturing facilities, data centers, some aerospace companies, and residential customers). Geopolitical challenges, conflicts, global changes in supply and demand, along with weather and climate, influence the price of LNG [2,44,82]. The implications of these price fluctuations have significant, albeit not yet fully understood, impacts on Louisiana's economy.

Louisiana stands at the center of a major economic transformation, one with profound global ties and intense local implications. While our documentation of spatially uneven economic conditions is hardly novel to the economic literature, what stands out here is the sheer magnitude of the unevenness. Hydraulic fracturing brought about a series of globe-changing dynamics, turning the United States from a major energy importer to a major energy exporter. The Gulf Coast was linked to new fossil energy supplies underway or recently completed, and more have been announced [40,41]. And yet, these changes also contributed to an economic downturn in Louisiana that led to localized job losses and economic downturns – and near-flat job growth statewide. Whether this is a common attribute of energy economies (*i.e.*, the resource curse) specific to hydraulic fracturing or linked to more modern changes in the nature of work, the transitions are far-reaching and of significant local and global importance.

Acknowledgement: The authors are very thankful to Anne Patton from With Bells On LLC for her excellent work preparing the manuscript.

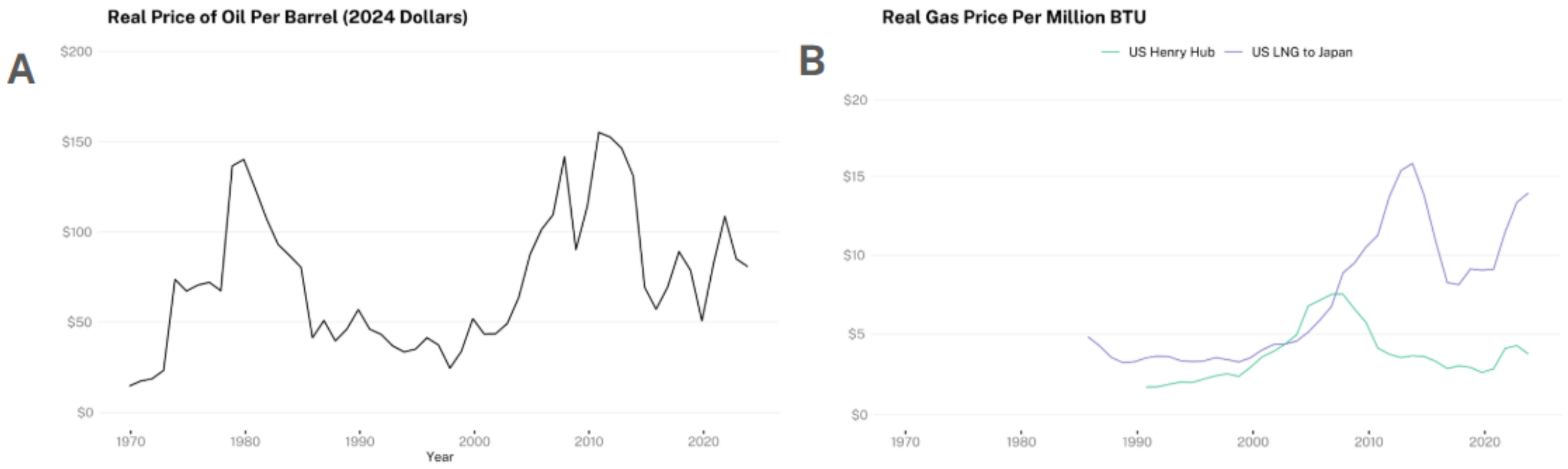

*Figure 1. Price of oil and natural gas. (A) Price of oil per barrel, normalized to 2024 US Dollars. (B) Price of natural gas in terms of dollars per BTU, including the price of LNG shipped to Japan (a common international metric for LNG), and the price of pipeline gas measured at the US Henry Hub (a standard metric for domestic US pricing). Data source: [6]*

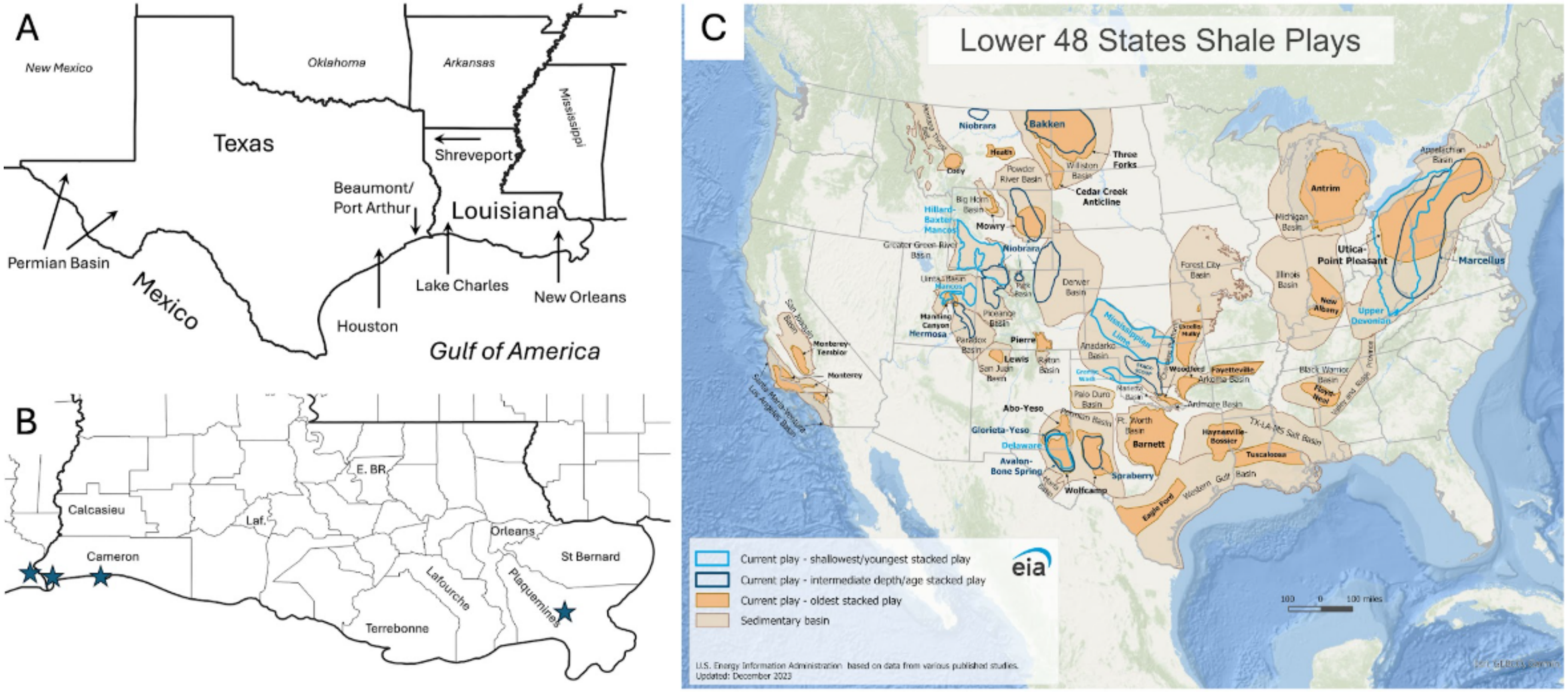

Figure 2 Regional and national maps. (A) Outline of the Gulf South region, noting the major states and cities. (B) Close-in view of south Louisiana, noting the location of major LNG facilities, and key parishes that are highlighted. (C) Shale plays (areas of extraction) in the lower 48 states. Source: (A) and (B) outlines are US Census TIGER files. (C) [15]

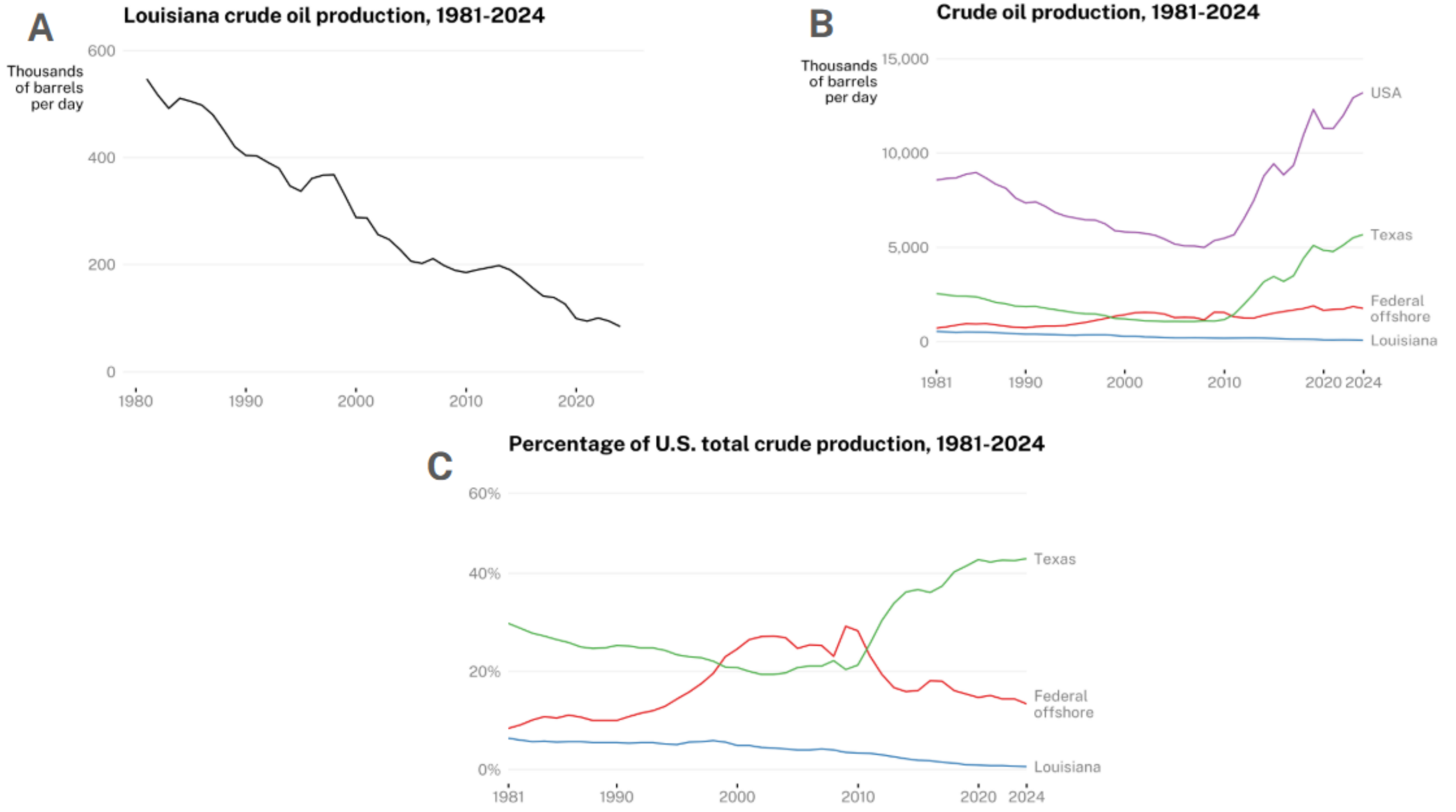

Figure 3.** Crude oil production in Louisiana and other key Gulf areas. (A) Louisiana production 1981-2024. (B) Production in Louisiana, Texas, the Federal Offshore Gulf of America, and the USA as a whole. (C) Percentage of US production in Texas, Louisiana and the Federal Offshore Gulf of America. (D) US Crude Oil export 1981-2023, and those with oil exports from PADD 3 – the district that includes Louisiana and Texas. Source: [20,21]

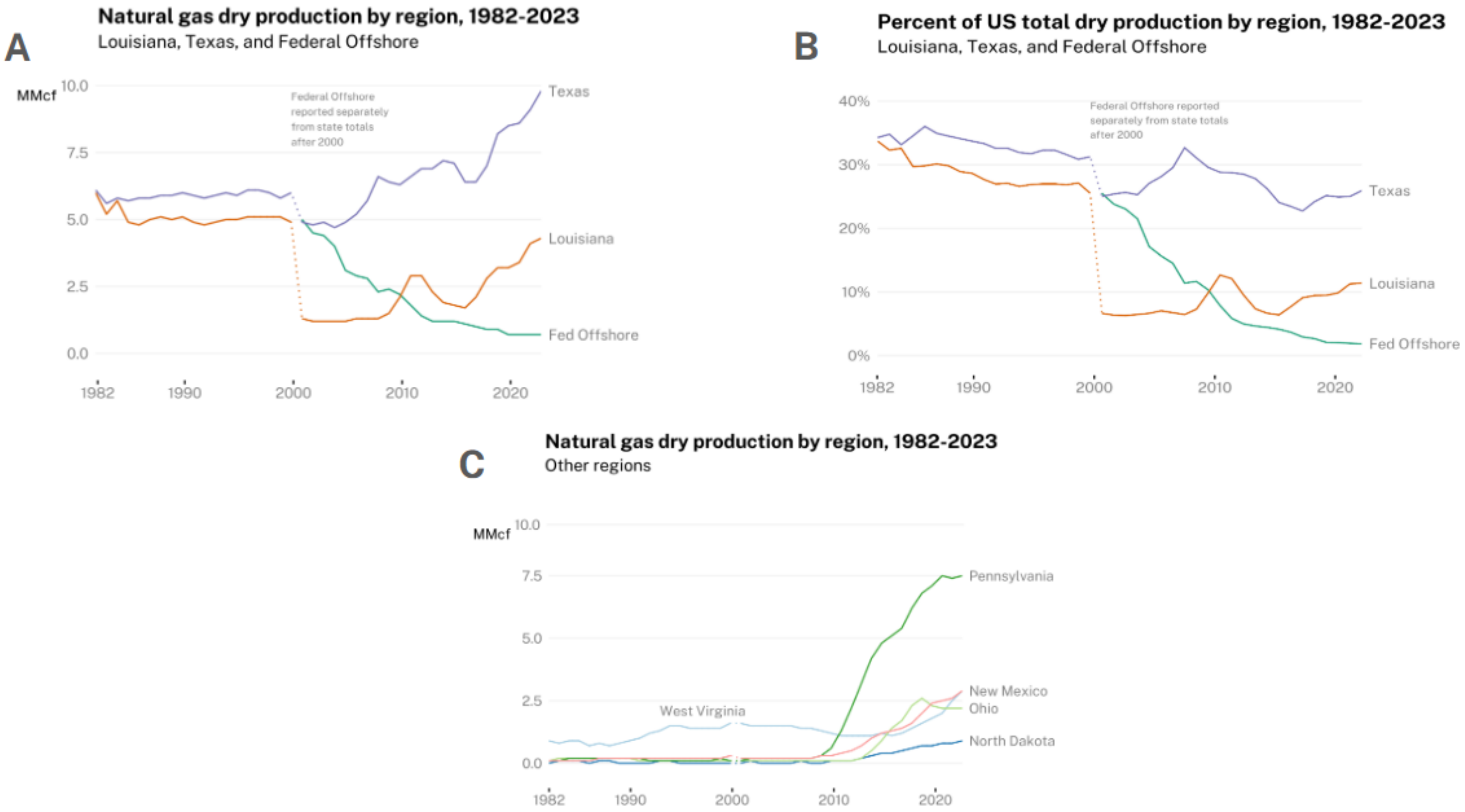

Figure 4 Natural gas production in the United States. (A) Volume produced in Texas, Louisiana, and the Federal Offshore Gulf of America. Note that after 2001, natural gas from the Federal Offshore Gulf was removed from Louisiana and Texas and placed in its own separate category. (B) The percentage of these 3 states as a fraction of the USA's total production. (C) Production in several other states with large increases that are linked to hydraulic fracturing. Source [22].

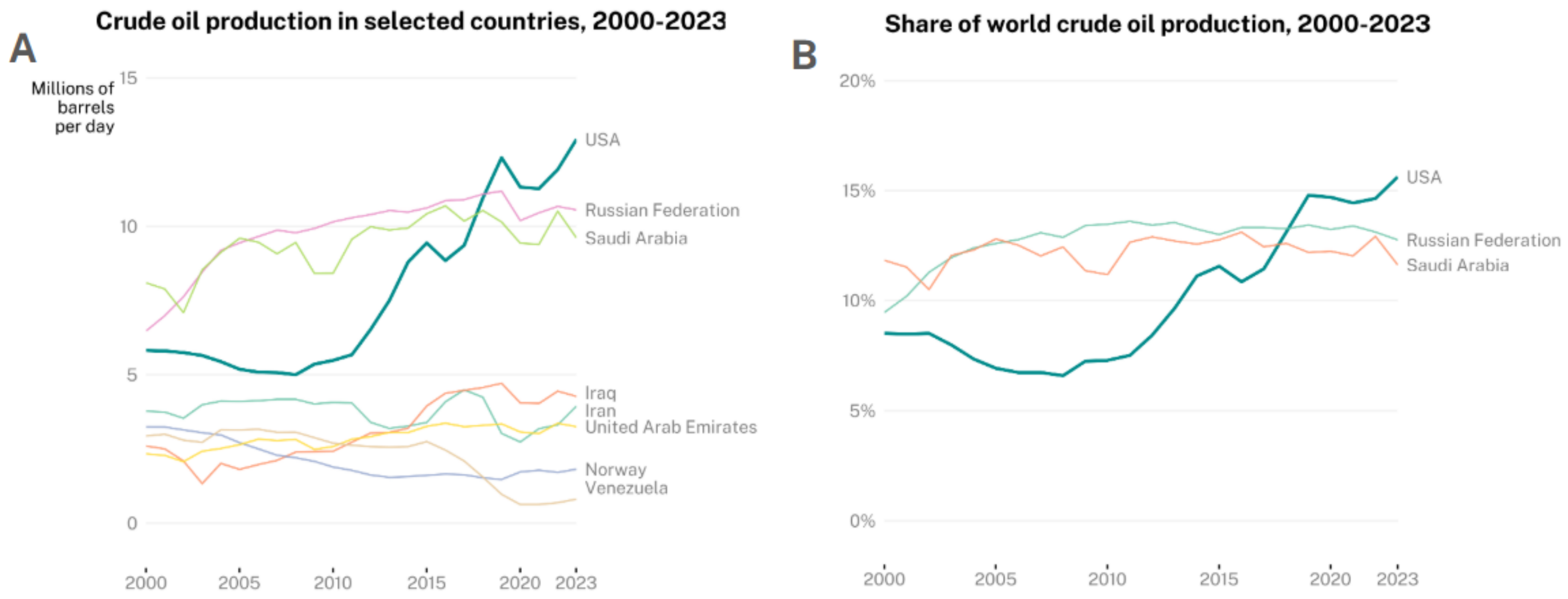

Figure 5 Crude oil production in some of the world’s largest producers. (A) Volumes of crude produced in the world's top 8 producers. (B) Crude oil production as a percentage of global production for the world's top three producers. Source [6].

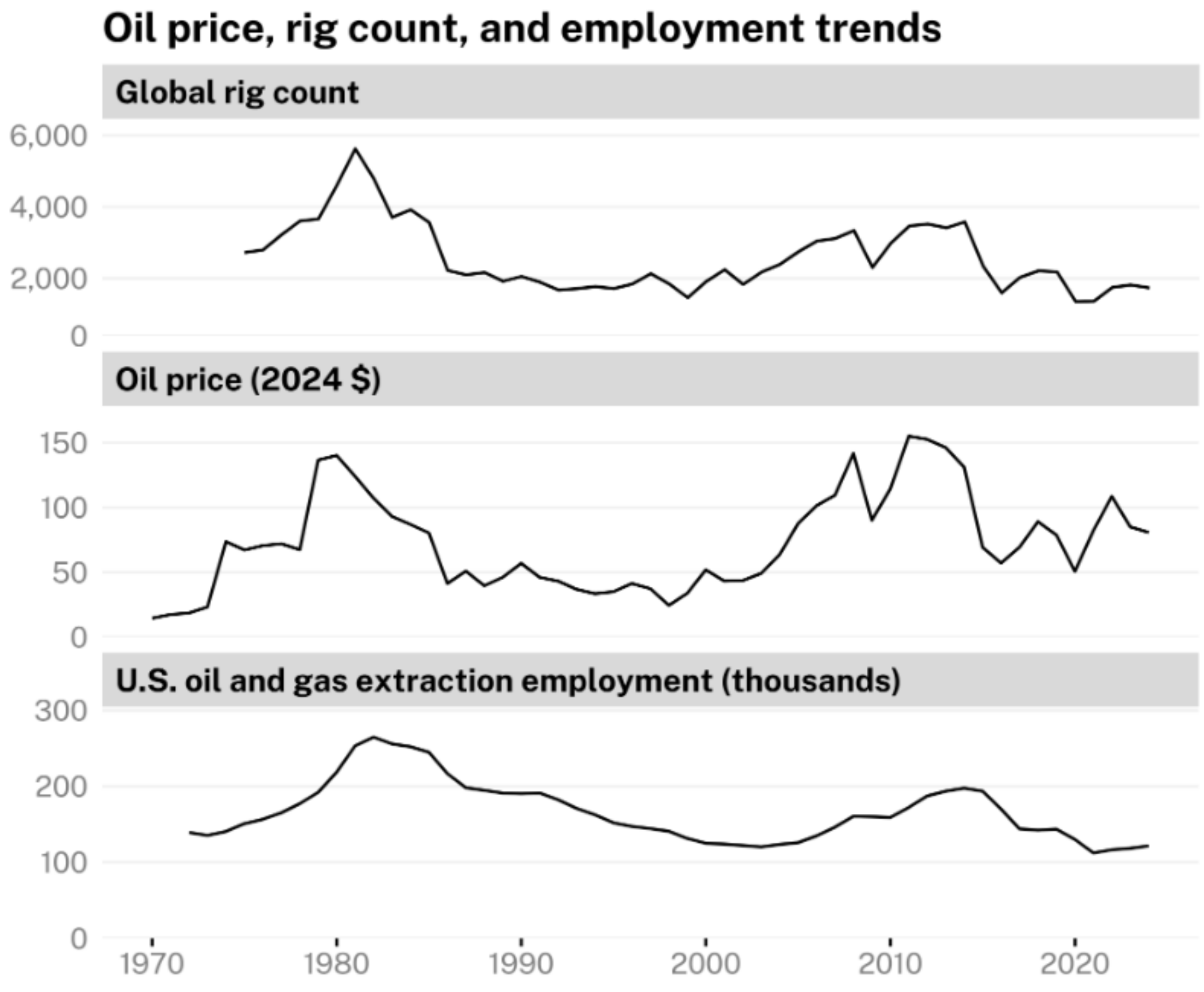

Figure 6 The global link between the price of oil, the global rig count, and oil and gas jobs in the United States. Note that the three fluctuate with approximately the same wave-like pattern, i.e. the super cycle. The global rig count is sourced from [26] the oil price data is sourced [6], using their method for determining prices in 2024 US Dollars, and the number of workers is derived from the US Quarterly Census of Employment and Wages.

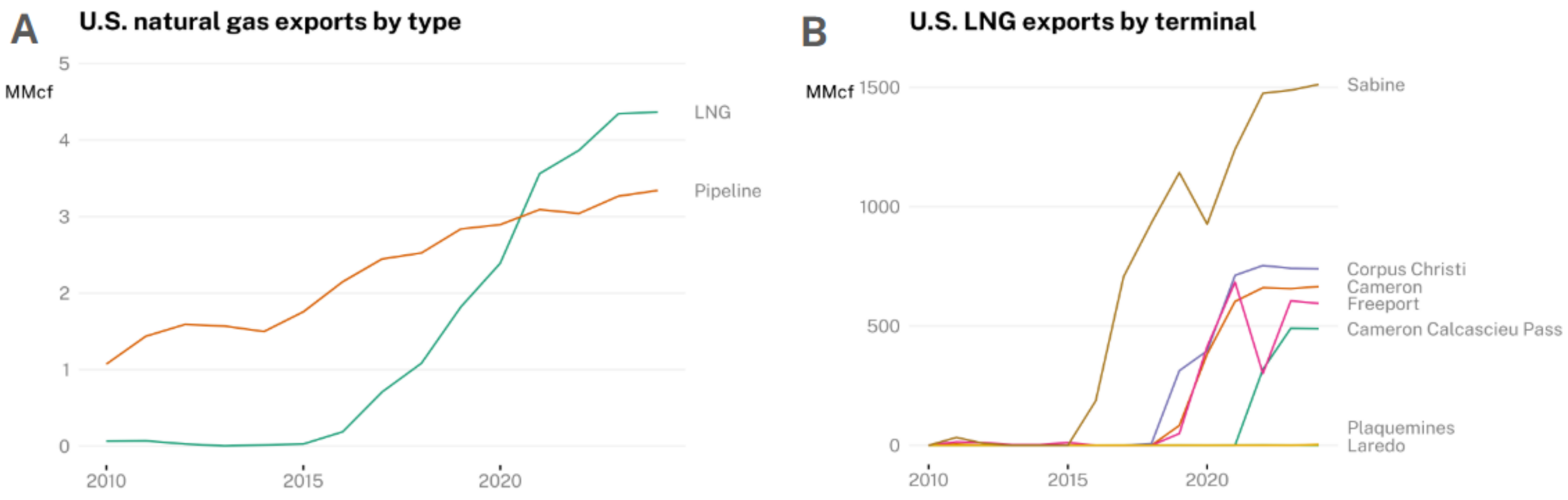

Figure 7 US natural gas exports. (A) Natural gas exports by type, LNG vs pipeline gas. (B) LNG exports by terminal. Source [6,59].

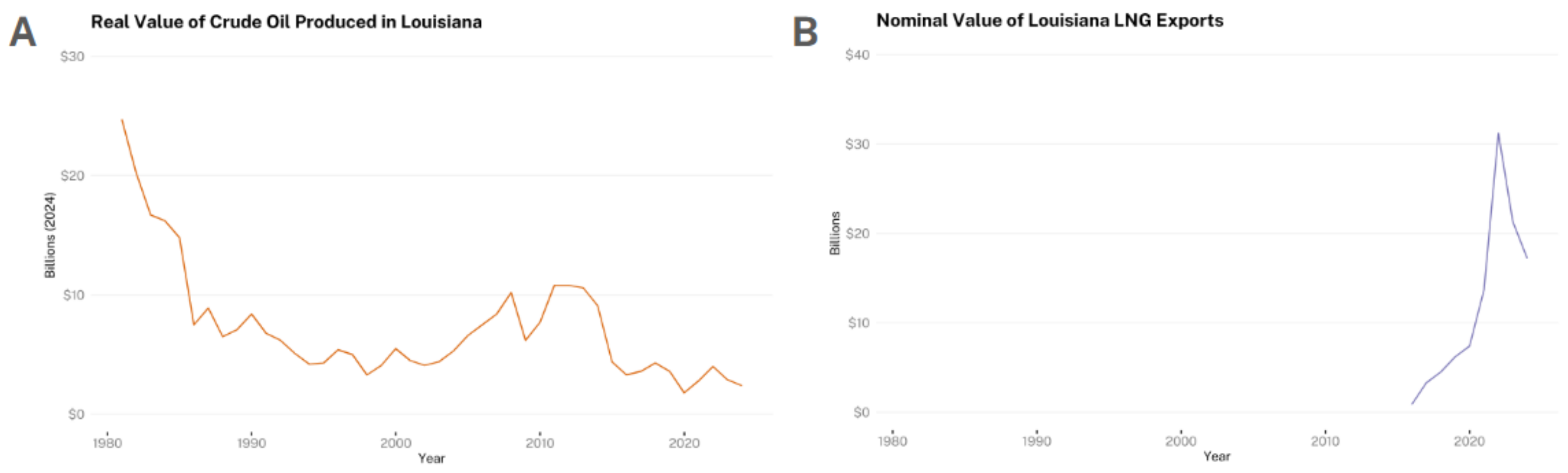

Figure 8 The relationship between oil production, (A) the price of oil (in 2024 dollars) and the total value of production within the state of Louisiana. (B) The value of LNG exported from Louisiana, as determined by the volume of exports and the price of LNG exported. Source: [6,32]

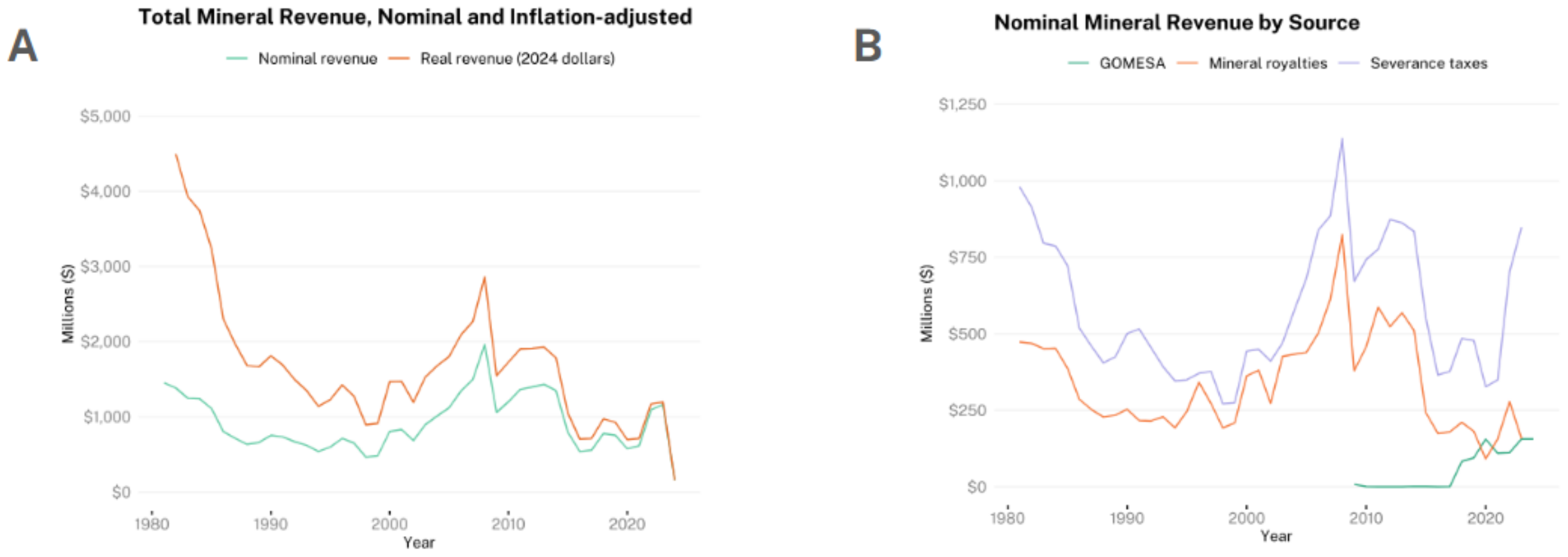

Figure 9 Sources of mineral revenue to the State of Louisiana. (A) Total mineral revenue, nominal and inflation-adjusted, (B) Nominal mineral revenue by source. Source: [58]

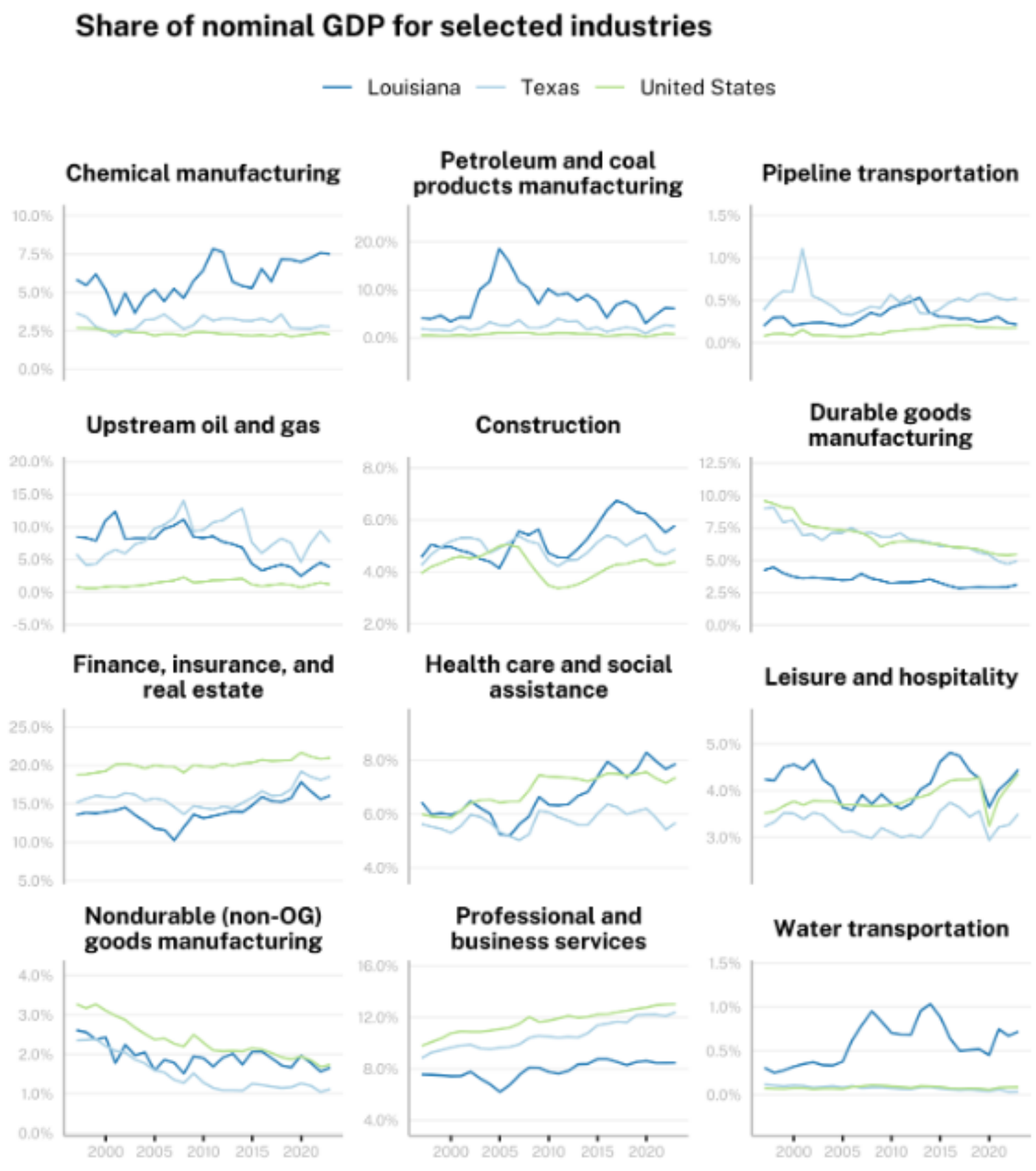

Figure 10 The share of nominal GDP for selected industries in Louisiana, Texas, and the United States. Source: US Bureau of Economic Analysis, GDP by industry in current dollars (SAGDP2).

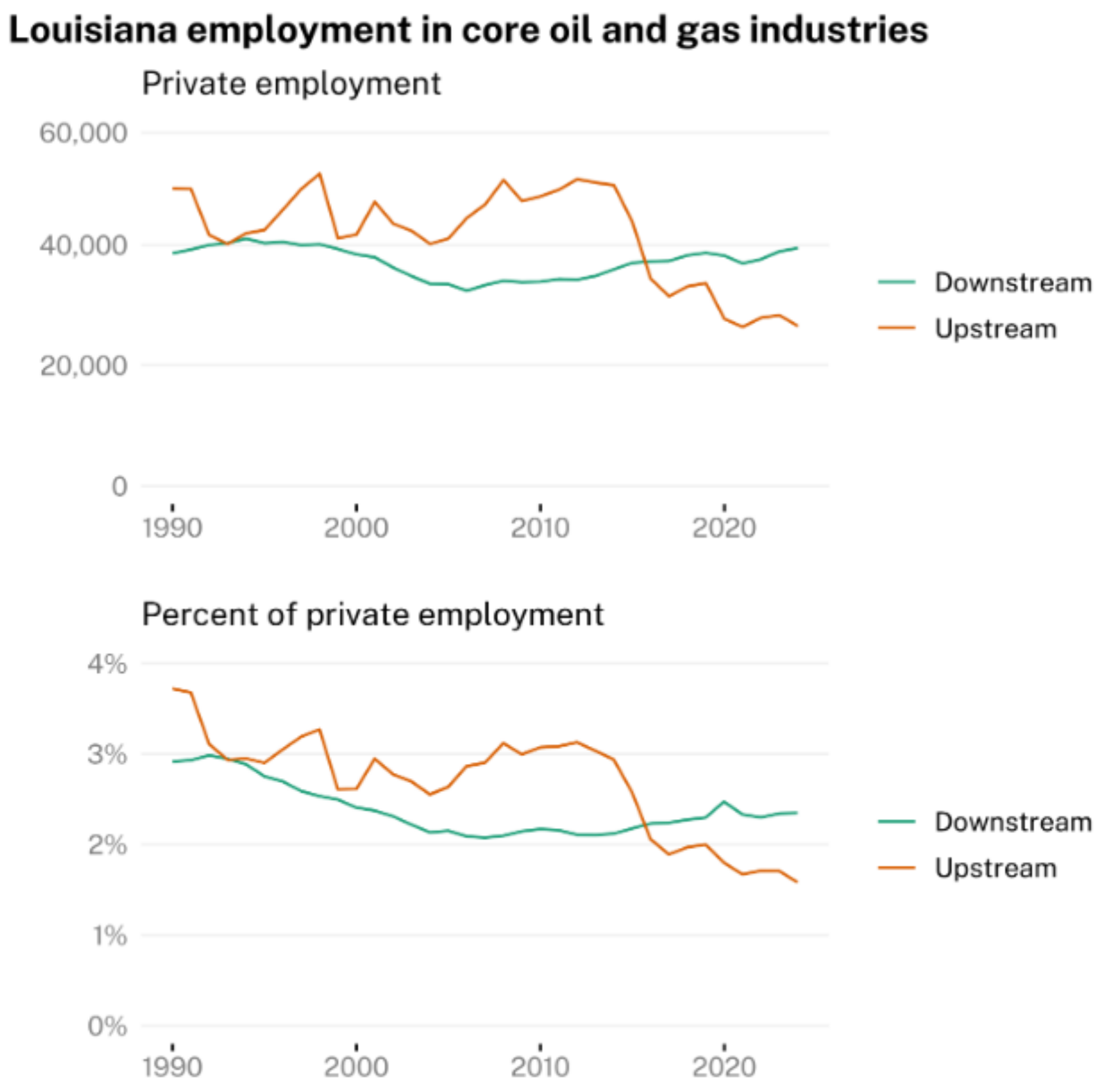

Figure 11 Private employment in Louisiana in upstream (production) and downstream (refining, chemical manufacturing etc.) sectors of the oil and gas industry: (A) Total employment. (B) Employment as a percentage of state total employment. Source: US Bureau of Labor Statistics, Quarterly Census of Employment and Wages (QCEW).

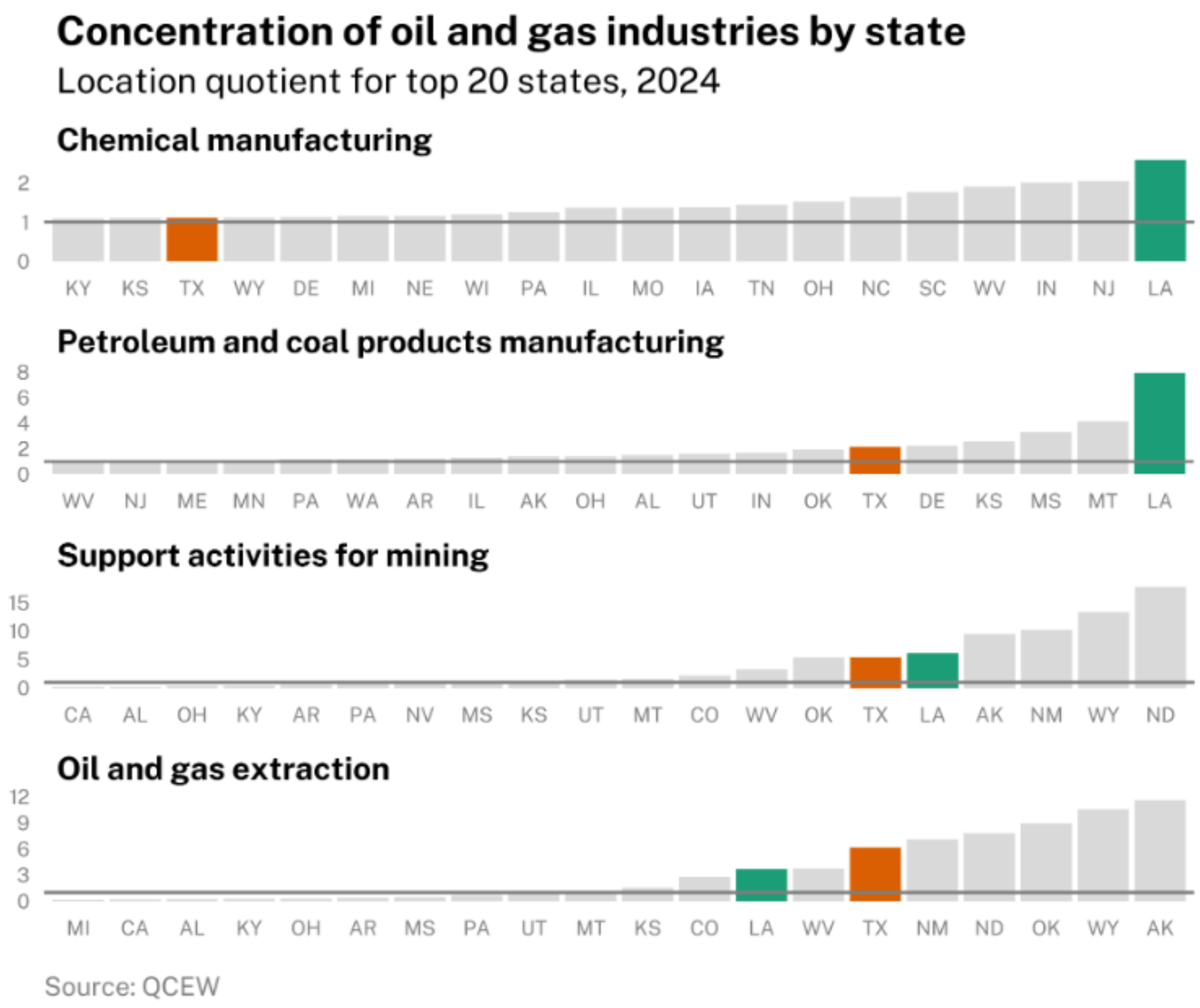

Figure 12 Concentration of key parts of the oil and gas industry in Louisiana as compared to other states using the location quotient (LQ). Source: US Bureau of Labor Statistics, Quarterly Census of Employment and Wages (QCEW).

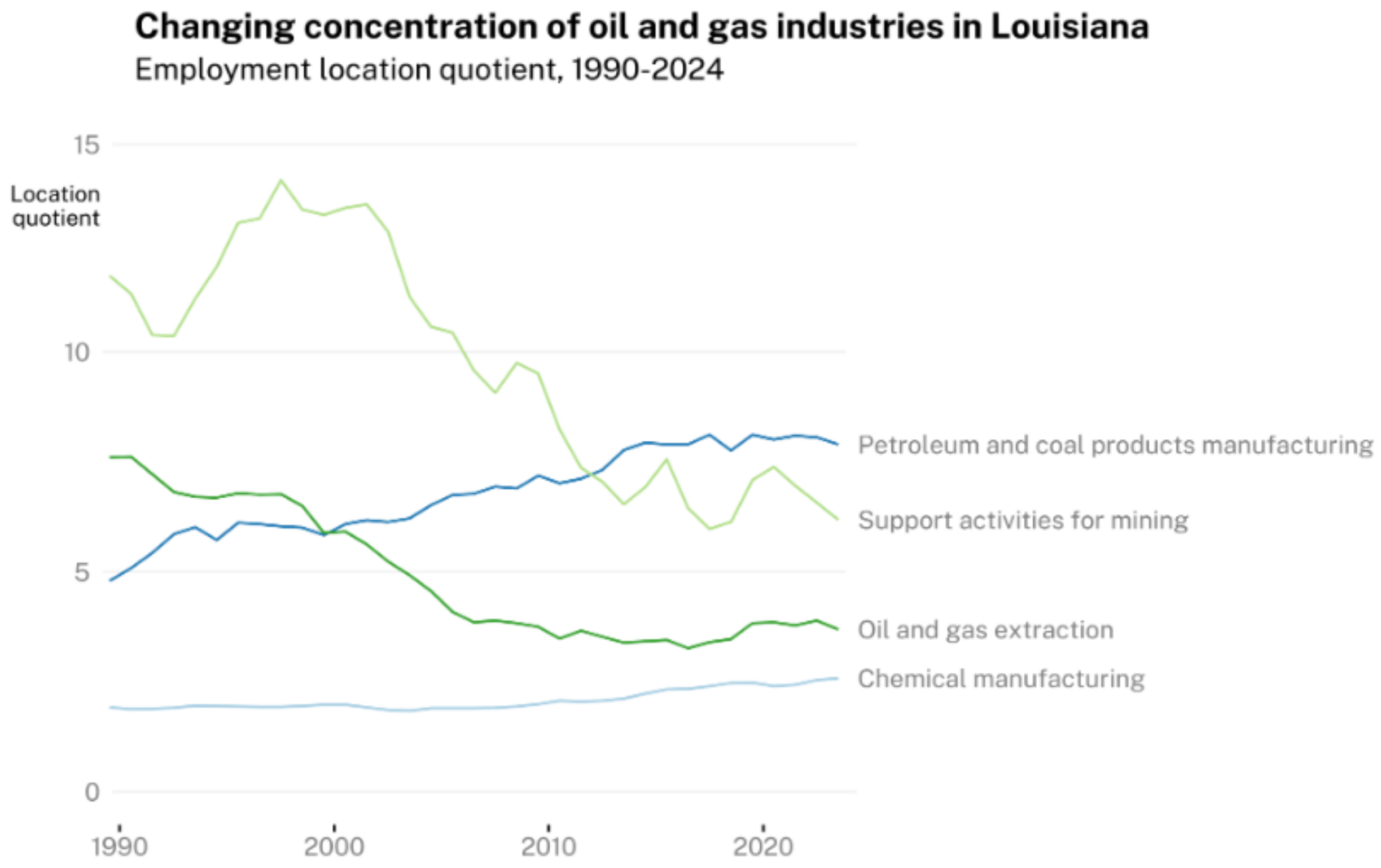

Figure 13 The changing nature of the location quotient of jobs in Louisiana. Note the downward trend in support activities for mining and the upward trend in petroleum and coal products manufacturing.

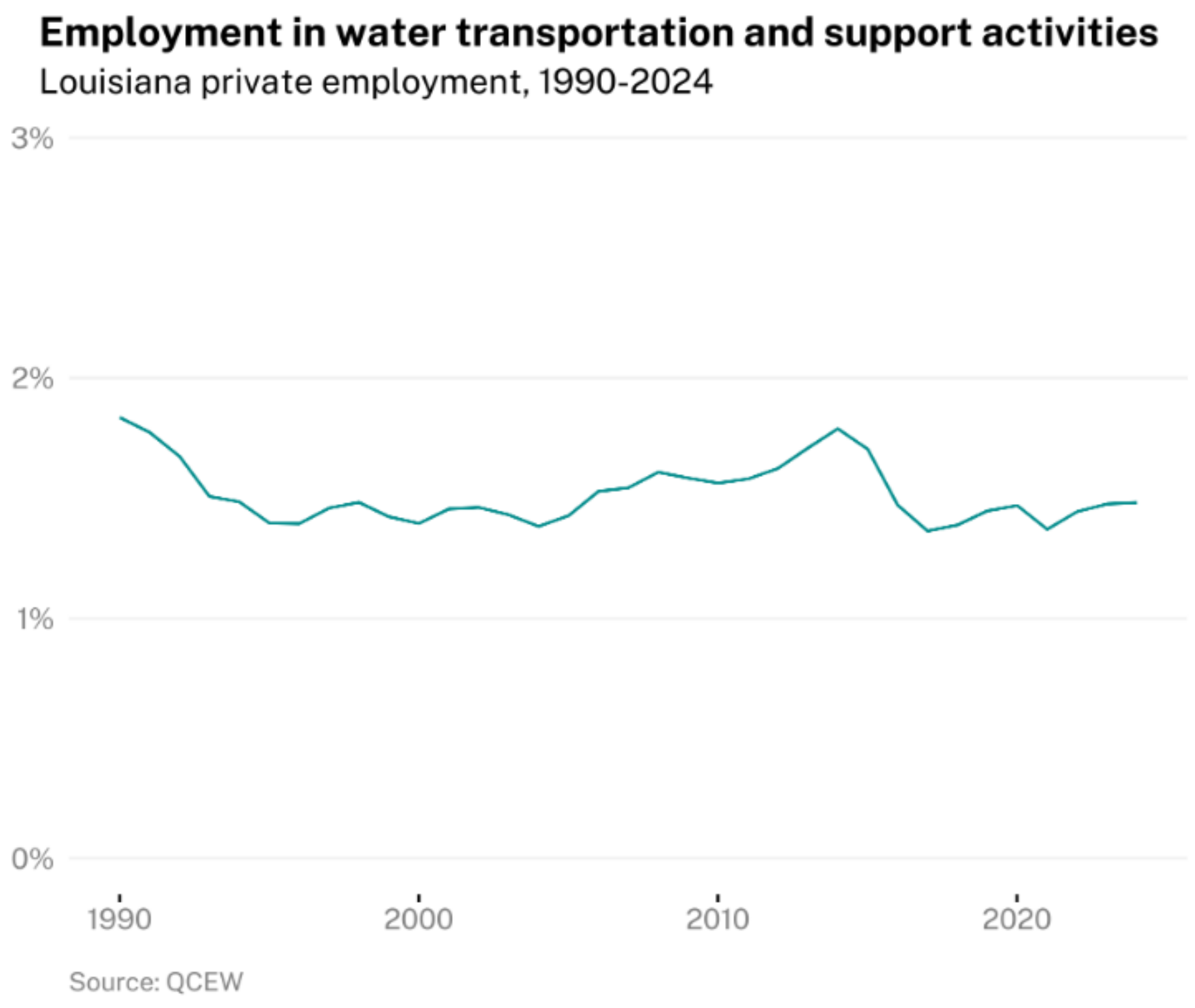

Figure 14 Employment in Louisiana in water transportation and support activities as a proportion of the total number of jobs in the state. Source: US Bureau of Labor Statistics, Quarterly Census of Employment and Wages (QCEW)

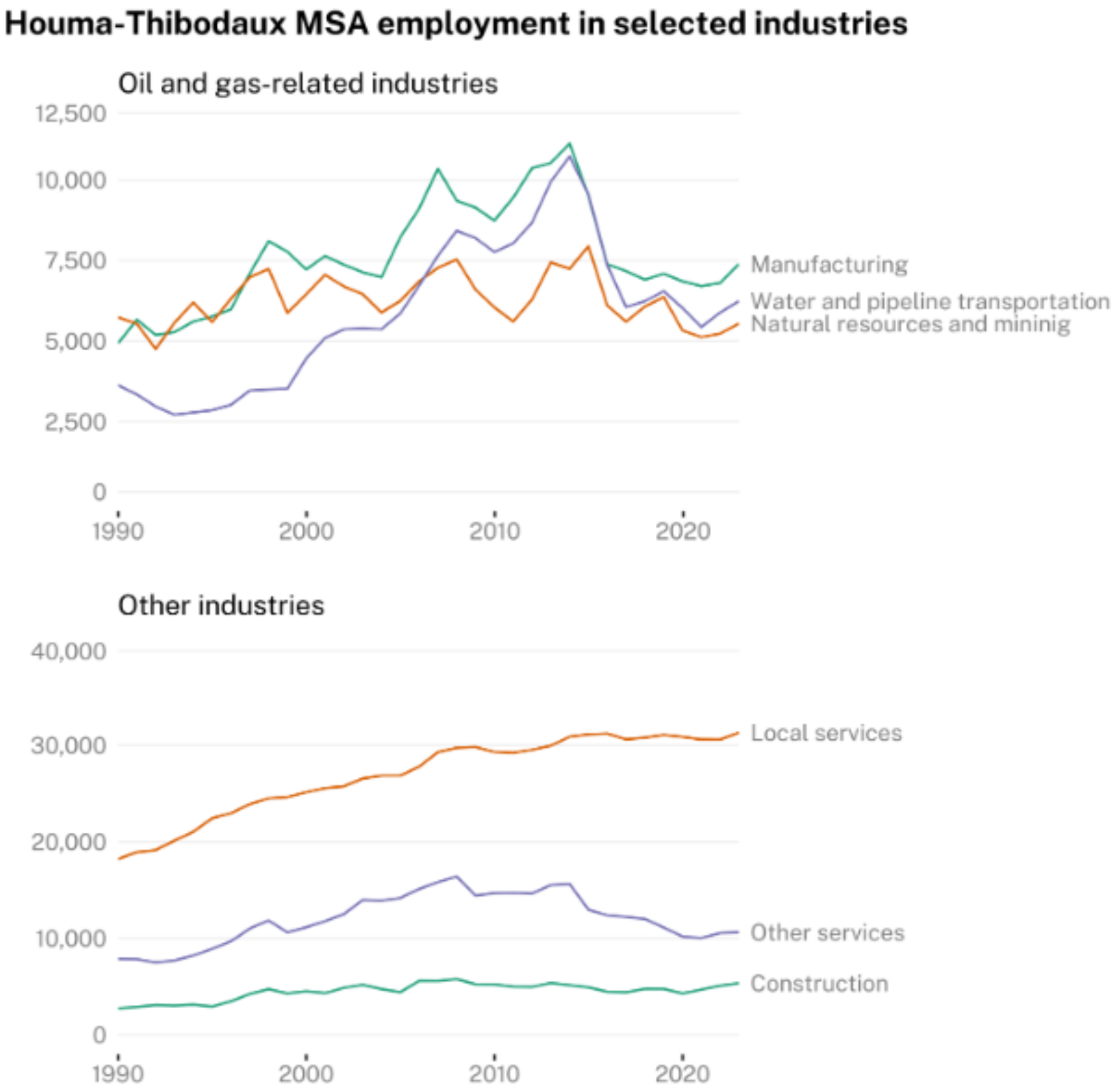

Figure 15 Employment in the Houma Thibodaux region in south-central Louisiana. The region has long been considered a hub for jobs in the offshore oil and gas industry. Source: US Bureau of Labor Statistics, Quarterly Census of Employment and Wages (QCEW).

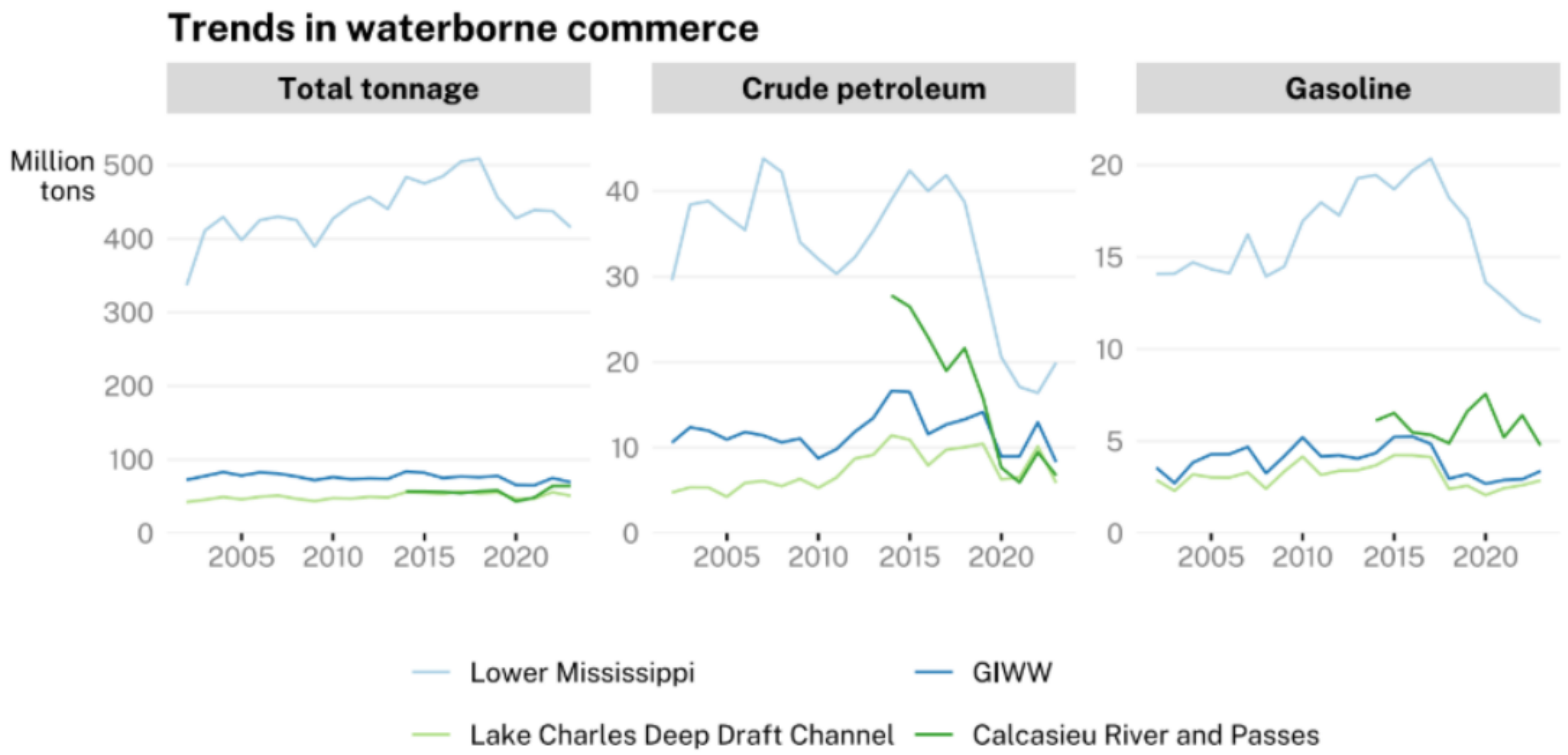

Figure 16 Trends in waterborne commerce in selected along key waterways in Louisiana. Trends are presented for the lower Mississippi River, the Gulf Intracoastal Waterway, and the Lake Charles Deep Draft. Source: [60]

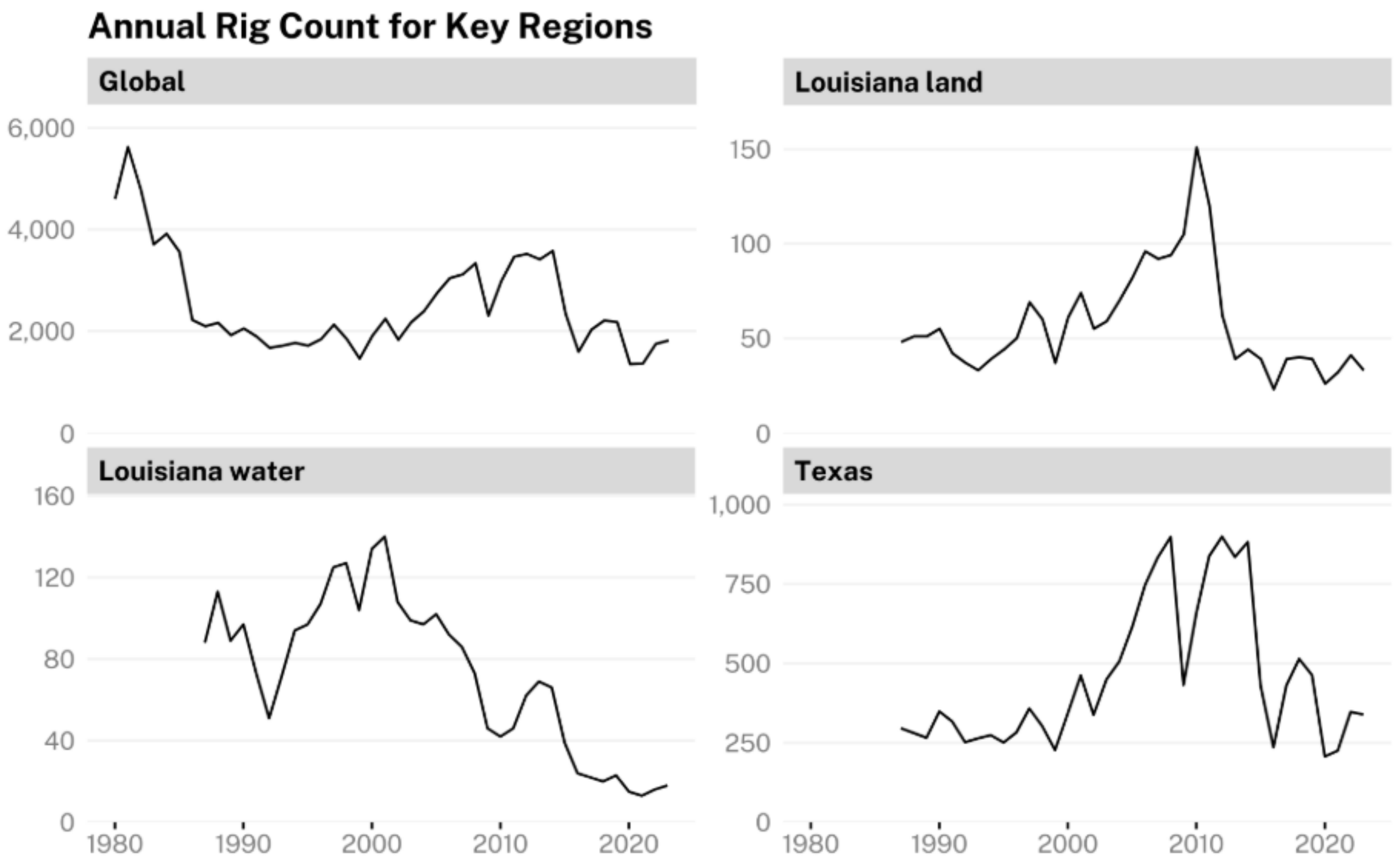

Figure 17 Number of drilling rigs in Louisiana, Texas, and the offshore regions of both states. (A) Total number in all locations. (B) Total number in Louisiana. (C) Total number in Texas. (D) Number of land-based and water-based rigs in Louisiana. Source: [61,62]

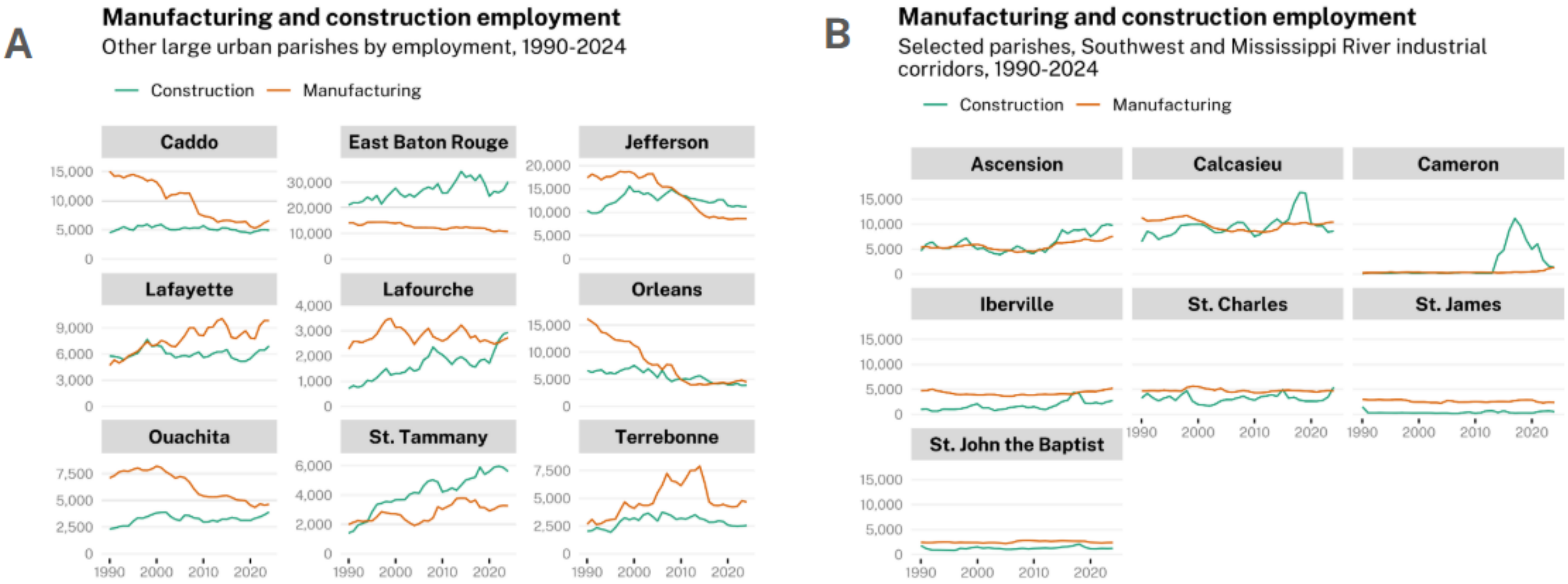

Figure 18 Employment in construction and employment in selected south Louisiana parishes: (A) Urban areas. (B) Parishes in southwest Louisiana and the Mississippi River industrial corridor. Source: US Bureau of Labor Statistics, Quarterly Census of Employment and Wages (QCEW).

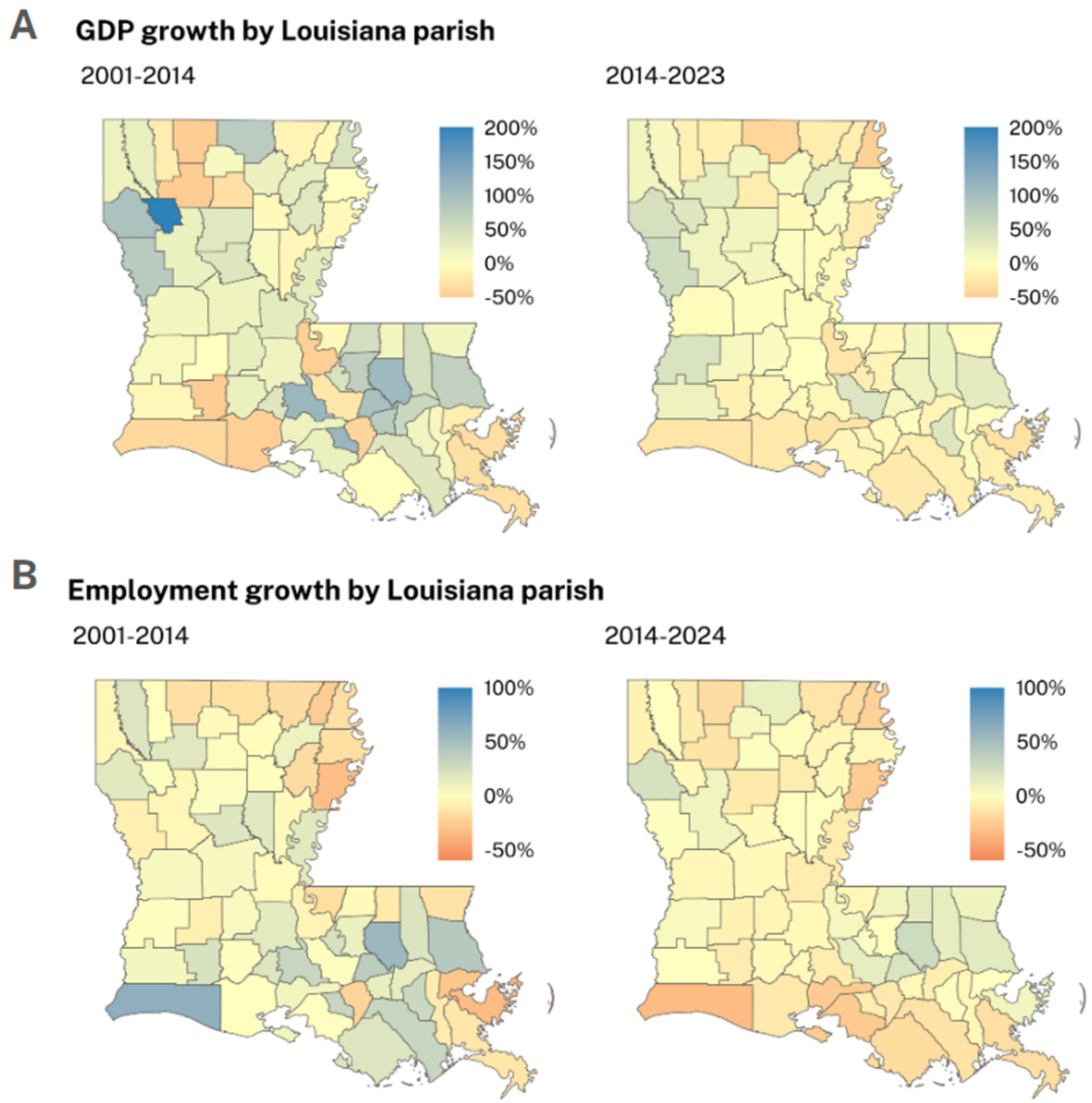

Figure 19. Maps of employment and GDP change in Louisiana's 64 parishes. Shown here are data for two time periods, 2001-2014 (before the oil price decline) and 2014 to 2023 (after the oil price decline). (A) GDP growth. (B) Employment growth. Sources: (A) US Bureau of Economic Analysis, Real Gross Domestic Product:], Retrieved from FRED, Federal Reserve Bank of St. Louis; (B) Source: US Bureau of Labor Statistics, Quarterly Census of Employment and Wages (QCEW).

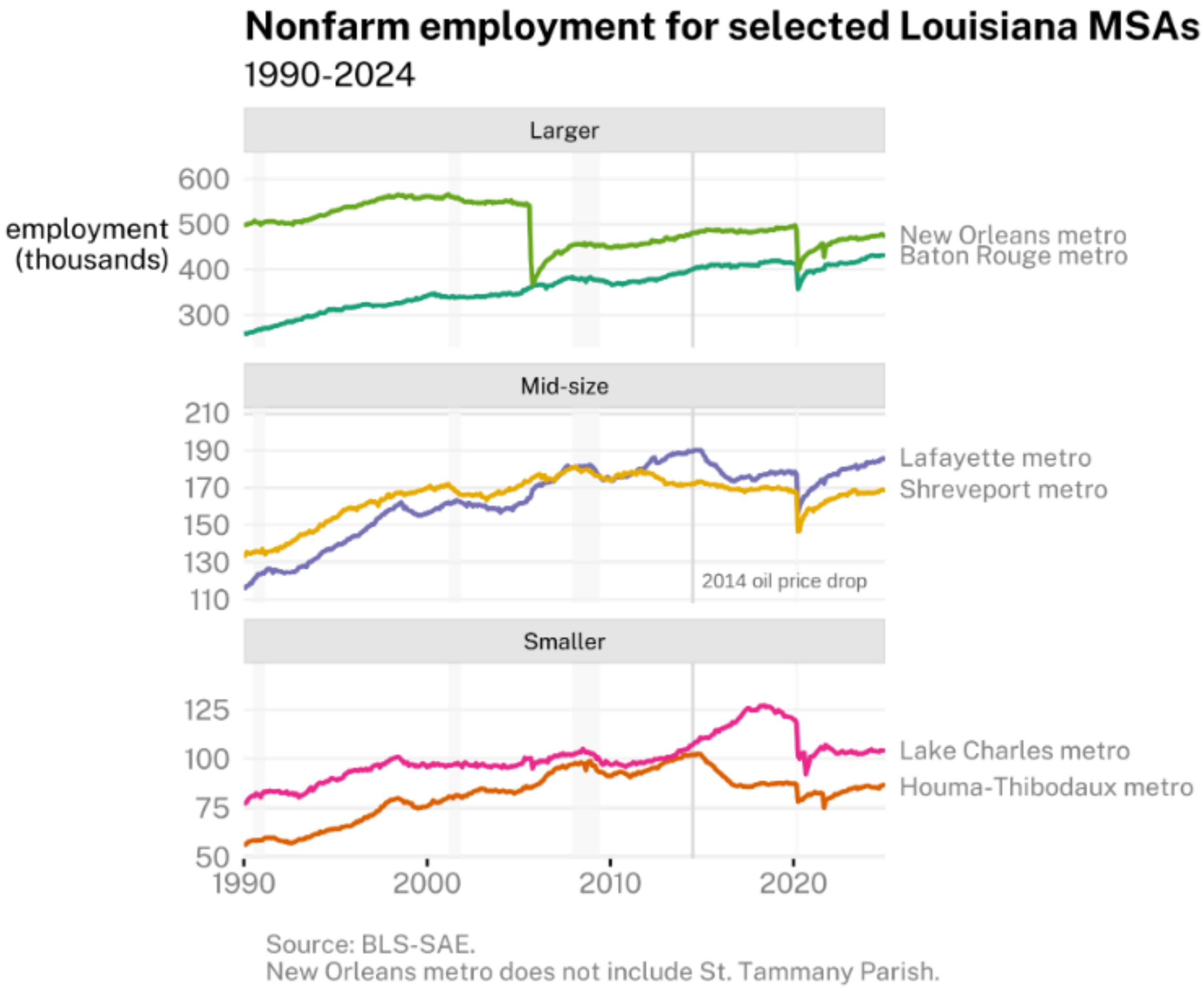

Figure 20 Non-farm employment for selected metropolitan statistical areas (MSAs) in Louisiana. Source: US Bureau of Labor Statistics' State and Area Employment.

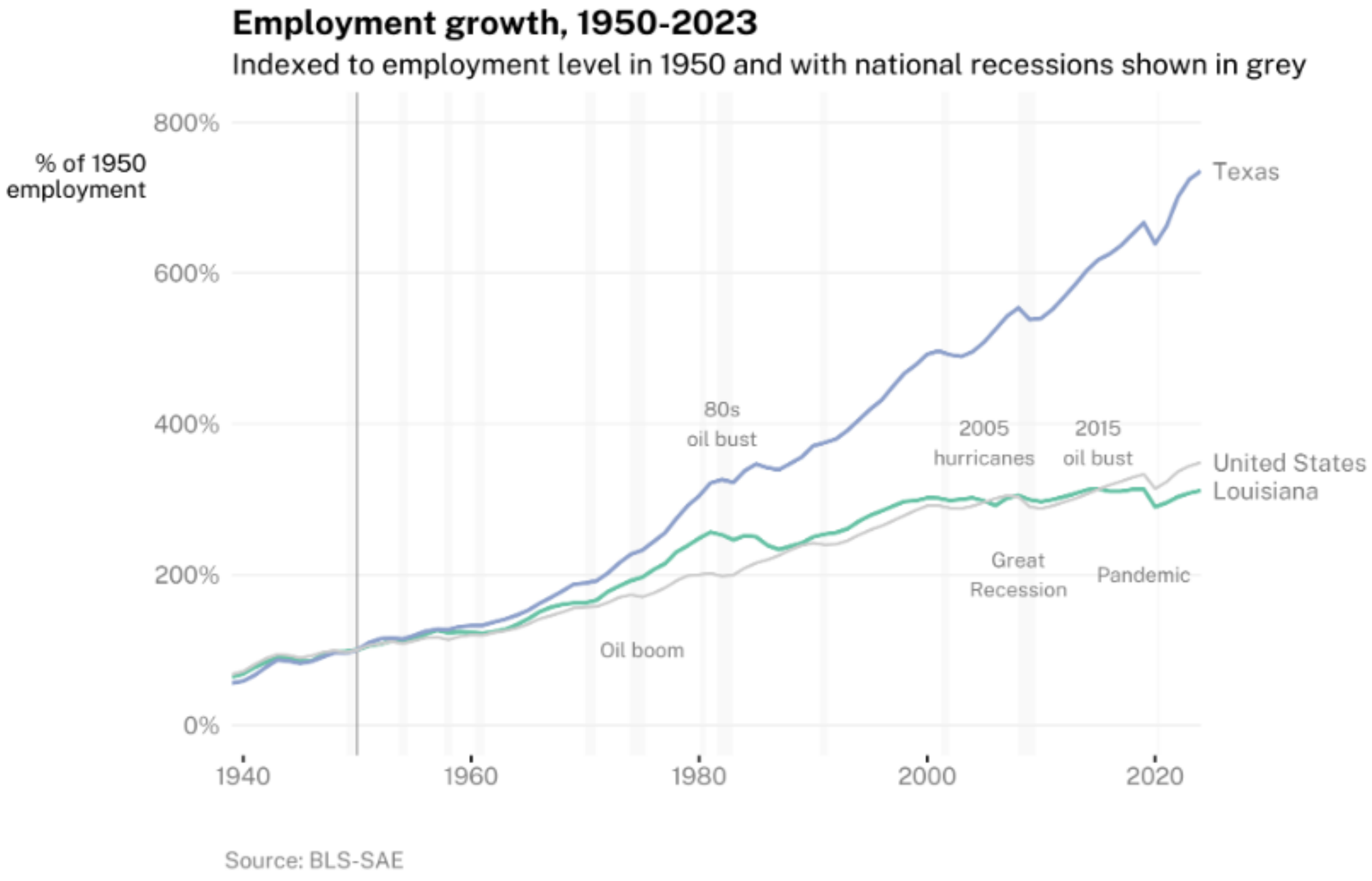

Figure 21 Employment in Louisiana, Texas and the United States, referenced to employment in 1950. Source: US Bureau of Labor Statistics: State and Area Employment

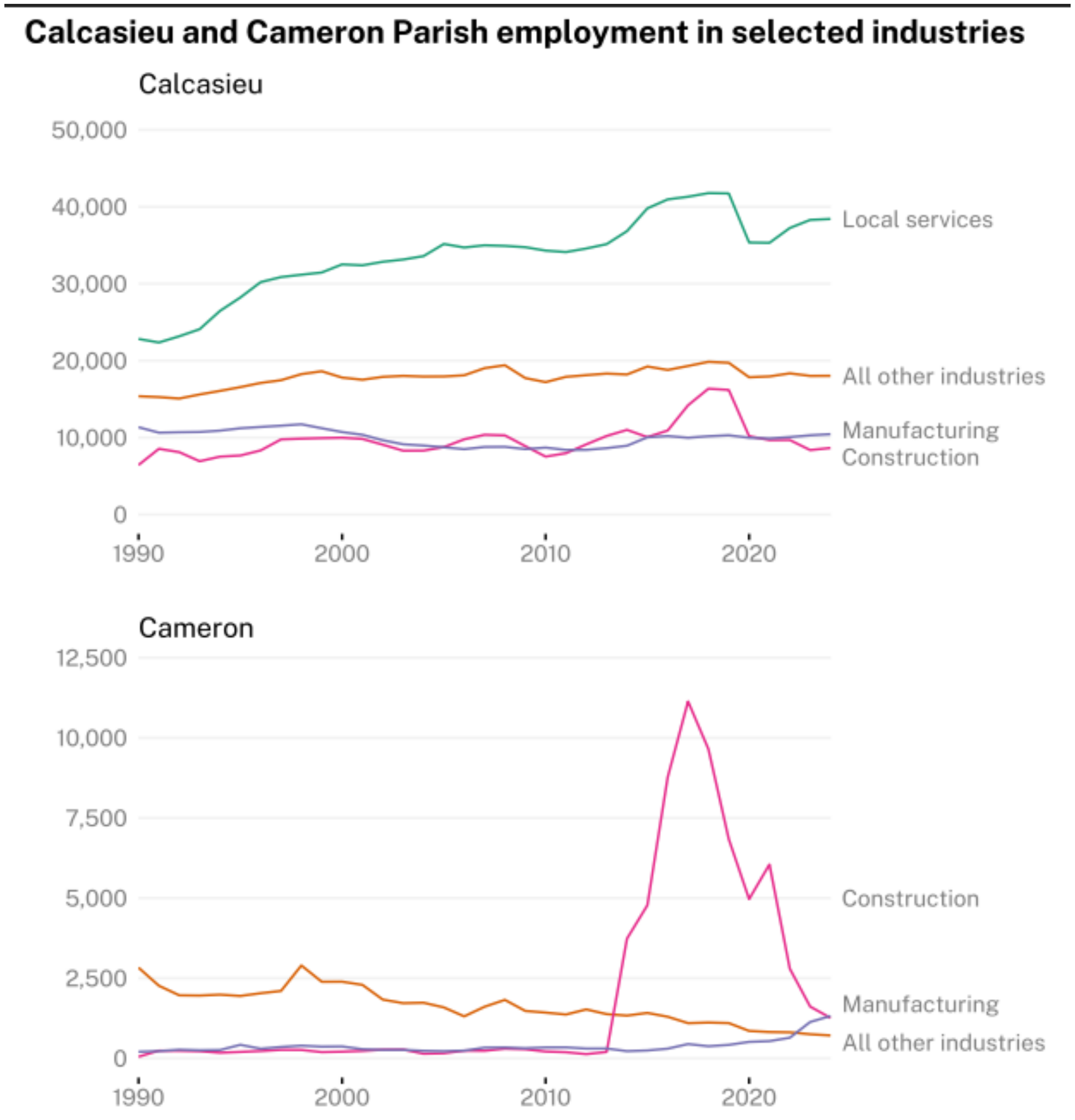

Figure 22 Employment in the Southwest corner of Louisiana concentration in Cameron and Calcasieu parishes. Source: US Bureau of Labor Statistics, Quarterly Census of Employment and Wages (QCEW)